\documentclass[12pt,preprint]{aastex}

\def\etal{{et al.}}
\def\ltsima{$\; \buildrel < \over \sim \;$}
\def\gtsima{$\; \buildrel > \over \sim \;$}
\def\lsim{\lower.5ex\hbox{\ltsima}}
\def\gsim{\lower.5ex\hbox{\gtsima}}
\def\lapp{\ifmmode\stackrel{<}{_{\sim}}\else$\stackrel{<}{_{\sim}}$\fi}
\def\gapp{\ifmmode\stackrel{>}{_{\sim}}\else$\stackrel{<}{_{\sim}}$\fi}

\newbox\grsign \setbox\grsign=\hbox{$>$} \newdimen\grdimen \grdimen=\ht\grsign
\newbox\simlessbox \newbox\simgreatbox
\setbox\simgreatbox=\hbox{\raise.5ex\hbox{$>$}\llap
     {\lower.5ex\hbox{$\sim$}}}\ht1=\grdimen\dp1=0pt
\setbox\simlessbox=\hbox{\raise.5ex\hbox{$<$}\llap
     {\lower.5ex\hbox{$\sim$}}}\ht2=\grdimen\dp2=0pt
\def\simgreater{\mathrel{\copy\simgreatbox}}
\def\simless{\mathrel{\copy\simlessbox}}
\newbox\simppropto
\setbox\simppropto=\hbox{\raise.5ex\hbox{$\sim$}\llap
     {\lower.5ex\hbox{$\propto$}}}\ht2=\grdimen\dp2=0pt

\shorttitle{GALEX paper I}
\shortauthors{Schiavon et al.}
 
\begin{document} 
\title{UV Properties of Galactic Globular Clusters with GALEX I. The
	Color-Magnitude Diagrams}

\author{Ricardo P. Schiavon}
\affil{Gemini Observatory, 670 N. A'Ohoku Place, Hilo, HI 96720, USA}
\email{rschiavon@gemini.edu}

\author{Emanuele Dalessandro}
\affil{Dipartimento di Astronomia, Universit\`a degli Studi di Bologna, via
	Ranzani 1, I-40127, Bologna, Italy}
\email{emanuele.dalessandr2@unibo.it}

\author{Sangmo T. Sohn}
\affil{Space Telescope Science Institute, 3700 San Matin Dr., Baltimore, MD
	21218, USA}
\email{tsohn@stsci.edu}

\author{Robert T. Rood\altaffilmark{1} \& Robert W. O'Connell}
\affil{Astronomy Department, University of Virginia, P.O. Box 400325,
	Charlottesville, VA 22904, USA}
\email{rwo@virginia.edu}

\author{Francesco R. Ferraro \& Barbara Lanzoni}
\affil{Dipartimento di Astronomia, Universit\`a degli Studi di Bologna, via
	Ranzani 1, I-40127, Bologna, Italy}
\email{francesco.ferraro3@unibo.it \& barbara.lanzoni@unibo.it}

\author{Giacomo Beccari}
\affil{ESO - European Southern Observatory, Karl-Swarzschild Str. 2, 
 D-85748 Garching bei Munchen, Germany}
\email{gbeccari@eso.org}

\author{Soo-Chang Rey}
\affil{Department of Astronomy and Space Science, 
Chungnam National University, Daejeon 305-764, Republic of Korea}
\email{screy@cnu.ac.kr}

\author{Jaehyon Rhee}
\affil{ Gemini Observatory, 670 N. A'ohoku Place, Hilo, HI 96720, USA}
\affil{Department of Physics, Purdue University,
525 Northwestern Avenue, West Lafayette, IN 47907, USA}
\email{jrhee@gemini.edu}

\author{R. Michael Rich}
\affil{Department of Physics and Astronomy, University of California, Los Angeles}
\email{rmr@astro.ucla.edu}

\author{Suk-Jin Yoon \& Young-Wook Lee}
\affil{Center for Galaxy Evolution Research and Department of Astronomy,
Yonsei University, Seoul, 120-749, Korea}
\email{sjyoon@galaxy.yonsei.ac.kr \& ywlee2@yonsei.ac.kr}
\pagebreak

\altaffiltext{1}{Deceased.}


\begin{abstract}

We present GALEX data for 44 Galactic globular clusters
obtained during 3 GALEX observing cycles between 2004 and 2008.
This is the largest homogeneous data set on the UV photometric
properties of Galactic globular clusters ever collected.  The sample
selection and photometric analysis are discussed, and color-magnitude
diagrams are presented.  The blue and intermediate-blue horizontal
branch is the dominant feature of the UV color-magnitude diagrams
of old Galactic globular clusters.  Our sample is large enough to
display the remarkable variety of horizontal branch shapes found
in old stellar populations.  Other stellar types that are obviously
detected are blue stragglers and post core-He burning stars.  The
main features of UV color-magnitude diagrams of Galactic globular
clusters are briefly discussed.  We establish the locus of post-core
He burning stars in the UV color-magnitude diagram and present a
catalog of candidate AGB-manqu\'e, post early-AGB, and post-AGB
stars within our cluster sample.

\pagebreak

{\it The authors dedicate this paper to the memory of co-author Bob
Rood, a pioneer in the theory of the evolution of low mass stars, and a
friend, who sadly passed away on 2 November 2011.}

\pagebreak
\end{abstract} 


\section{Introduction} \label{intro}

It is fair to say that the last frontier of our growing understanding
of the physics of old stellar populations resides in the ultra-violet
(UV). The behavior of old stellar populations in the UV has puzzled
astronomers for almost four decades now, and in spite of major
recent progress, there are still important gaps in our understanding
of the nature of the stars that dominate the integrated light of
old stellar populations in the UV---particularly the far-UV \citep[FUV,
e.g.,][]{fe98,oc99,mo01,ca09}.  These include the extreme horizontal
branch (EHB) and Blue-Hook (BHk) stars, at the hot and visually
faint end of the horizontal branch (HB), and the short-lived but
more luminous supra-HB and post-asymptotic-giant-branch (PAGB)
stars.  Another population whose nature is still not entirely well
understood is that of blue straggler stars, which at the characteristic
ages of Galactic globular clusters (GCs) are not hot enough to
contribute substantially to the integrated FUV light, but are an
important source of near-UV light \citep[NUV, e.g.,][]{fe01,fe03},
and are in some cases detectable in integrated light longward of
3400~${\rm\AA}$ \citep[e.g.,][]{tr05,s07}.

The UV properties of old stellar populations have been a subject
of intense scrutiny ever since the discovery of the ``UV-upturn''
of early-type galaxies (Code 1969). While it has become clear in
the past decade or so that EHBs are responsible for most of the
``excess'' UV emission observed in old stellar populations
\citep[e.g.,][]{gr90,gr99,do95,oc99,br01}, our understanding of the
physics underlying the structure and evolution of such stars is
still plagued by theoretical uncertainties. Undeniably, uncertainties
are partly due to the absence of an accurate, comprehensive,
statistically representative, homogeneous dataset presenting the
colors and magnitudes of the stars responsible for the UV emission
in Galactic GCs---in spite of painstaking observational efforts by
a number of groups \citep[for reviews see, e.g.,][]{oc99,mo01}.
A database of that kind would also have important applications for
studies of extra-galactic stellar populations, as it could be used
to unveil correlations between features in the color-magnitude
diagrams (CMDs) of stellar populations and their integrated properties.
Such correlations can help understand the nature of distant systems,
for which only integrated properties are available.  In particular,
direct comparisons between integrated UV properties of Galactic
and extra-galactic GCs \citep[e.g.,][]{so06,re07,re09}
can lend insights on the stellar population content of those systems
(Dalessandro \etal\ 2012, in preparation, hereafter Paper II).

With this motivation in mind, we decided to use the {\it Galaxy Evolution
Explorer} (GALEX) to undertake the largest ever systematic and
homogeneous census of the UV properties of Galactic GCs. Data were
collected for 44 clusters in 3 GALEX cycles, from which UV CMDs and
integrated colors were obtained.  This paper discusses the sample
selection and the photometric analysis of the data.  Some of the data
have been used in combination with HST and ground-based observations
for multi-band photometric investigations of the stellar populations
of NGC~1904 \citep{la07} and M~2 \citep{da09}, from the innermost
regions to the extreme outskirts of those clusters. Paper II describes
the derivation of integrated magnitudes for this cluster sample,
and presents an analysis of correlations between integrated magnitudes and
colors and global cluster properties.  Paper III (Rood \etal 2012,
in preparation) introduces a new classification scheme of the HBs
of Galactic GCs, based on their UV morphologies.

This paper is laid out as follows.  Section~\ref{obsdata} describes
the sample selection, observations, data reduction and analysis.
In Section~\ref{cmds} the CMDs are presented.  A description of our
new catalog of post He-core-burning star candidates is presented in
Section~\ref{uvbright}.  Our conclusions are summarized in
Section~\ref{epilogue}

\section{Observations and Data Analysis} \label{obsdata}

GALEX is a 50~cm orbiting UV telescope launched in April 2003.  GALEX
has a circular field of view of $\sim 1.2\deg$ diameter and, in
imaging mode, a dichroic beam splitter allows it to collect data in
two simultaneous channels, in $FUV$ and $NUV$ bands, corresponding to
$\lambda=1350-1780$ and $1770-2730$~${\rm\AA}$ ($\lambda_{eff}$
= 1516 and 2267~${\rm\AA}$) and with a spatial resolution of 
$\sim$4.5${\arcsec}$
and 5.5${\arcsec}$, respectively.  GALEX detectors consist of two
stacks of three large format microplate channels and associated
electronics inserted in sealed tubes.  The $NUV$ and $FUV$ detectors
differ mostly in terms of the photocatode material (CsI in the case of
$FUV$, and Cs$_2$Te in the case of $NUV$) and the windows (MgF$_2$ for
$FUV$, fused silica for $NUV$).  The GALEX detectors record lists of
time-ordered photon positions and pulse heights, and these are
pipeline-processed on the ground for image reconstruction.  The
resulting images have pixel scales of 1.5${\rm \,arcsec\,pixel^{-1}}$
in both $FUV$ and $NUV$.  Both detectors can be damaged by high global and
per/pixel count rates, which prevents targeting very (UV-) bright
stars and the low Galactic latitude regions, due to their high UV
background.  For more details, see \cite{mo05}, \cite{mo07} or the
GALEX instrument overview at {\tt
http://galexgi.gsfc.nasa.gov/docs/galex/Documents/MissionOverview.html}.

\subsection{Sample and Observations} \label{sampleobs}

The data presented in this paper were primarily collected under
GALEX GI programs \#056 and \#099 (PI R.\ Schiavon) in Cycles 1 and
4, respectively.  The target selection was performed with an eye
towards spanning a wide range in metallicity and HB morphology.
Limitations, however, were imposed by target magnitudes and the
safety of GALEX's UV detectors, so that very distant, heavily
extinguished, and low Galactic latitude clusters could not be
included in the sample.  As a result, a number of interesting
clusters, particularly metal-rich ones at low Galactic latitudes,
were not observed, because the high UV background in these regions
could potentially harm the GALEX detectors.  Clusters with very
UV-bright stars within the GALEX field of view also could not be
observed (most notably $\omega$ Cen and NGC~6752) and in some cases
the pointing had to be adjusted in order to exclude such stars from
the field of view.  The target list for Cycle 1 totalled 25 Galactic
GCs and was composed primarily of clusters for which EHB stars could
be detected in both $FUV$ and $NUV$ bands in a single GALEX orbit,
with typical exposure times of 1,500 s.

For Cycle 4, our strategy entailed deeper exposures on a smaller
sample of 15 clusters, with a focus on expanding coverage towards
higher metallicity and younger age, while including clusters of
known extra-galactic origin, such as Arp 2 and Terzan 8
\citep[e.g.][]{lm10}.  We also took advantage of the relaxation of
the UV-brightness constraints dictated by detector-safety considerations
in order to obtain data for metal-rich Galactic GCs at relatively low
Galactic latitude, such as NGC~6342 and NGC~6356.  For the latter,
as well as for very distant Galactic GCs (NGC~2419, Terzan 8, Arp 2, IC
4499) we originally had little hope of obtaining good quality CMDs,
and just aimed at measuring reliable integrated colors (but see
discussion in Section~\ref{cmds}).  For one cluster (NGC~6273)
$NUV$ data were not collected.

Finally, we further include data for 6 out of 8 Galactic GCs from
Cycle 3 GI program \#075 (PI S.T. Sohn), which aimed at measuring
reliably UV fluxes of extreme HB stars in the program clusters to
test the helium-rich hypothesis for the production of EHB
stars \citep[e.g.,][]{le05}.  We plan to present the results of this
analysis in a forthcoming paper (Paper IV, Sohn \etal\ 2011, in
preparation).  For the Cycle 3 program we selected clusters that
exhibit extended HB blue tails in their optical CMDs.

\begin{figure}[!hp]
\begin{center}
\includegraphics[scale=0.7]{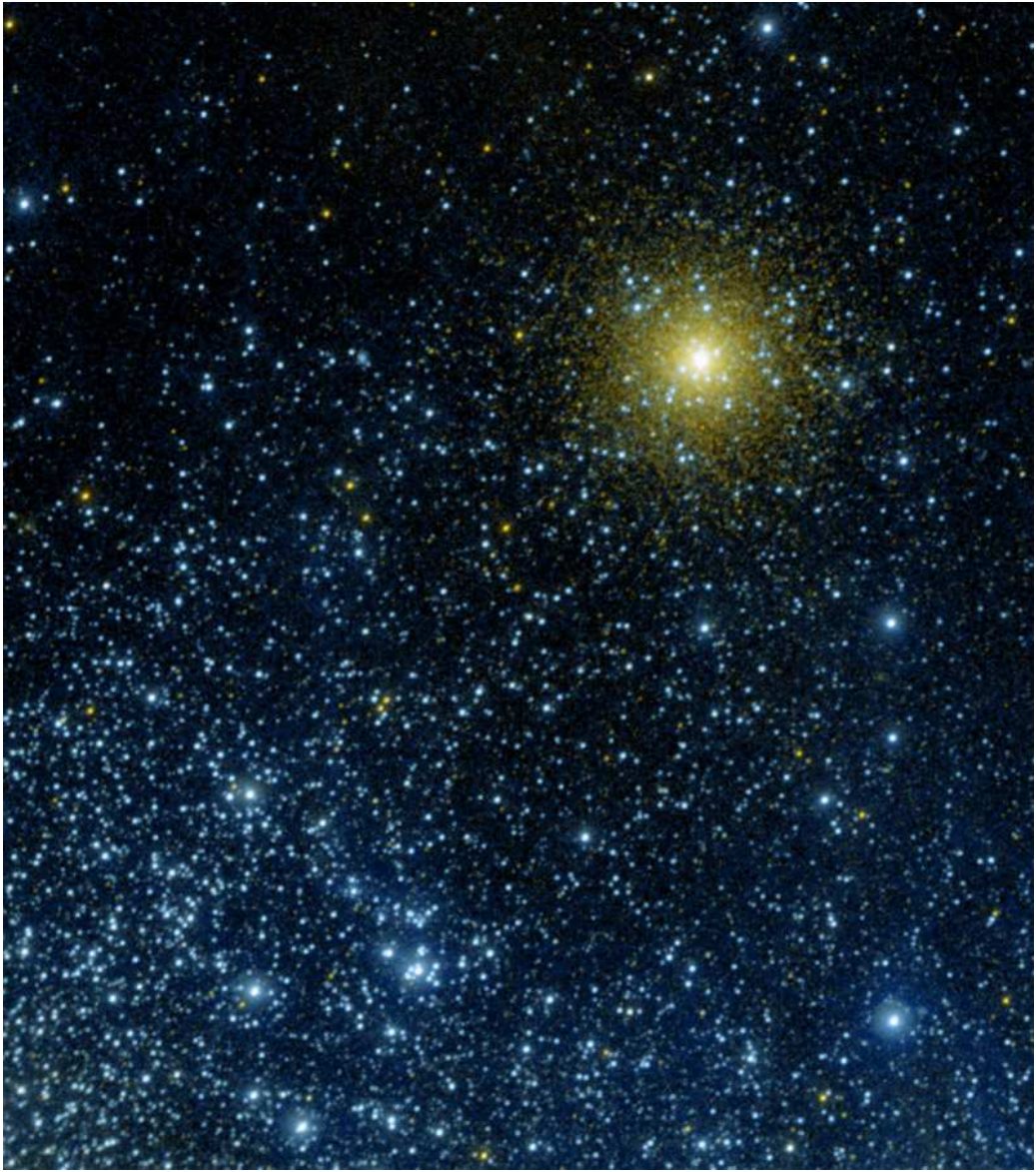}
\caption{False color picture of one of the fields targeted in our survey.
North is up, East is left, and the vertical size is $\sim$ 37\arcmin.  
There is a population of very blue stars covering the entire field, with a
higher density towards the SE.  These are main sequence stars belonging to
the Small Magellanic Cloud.  The cluster located on the upper right of the
picture is NGC~362.  Blue horizontal branch stars in NGC~362 appear as
white colored objects within a few arc minutes from the cluster center.
}
\label{colorpic}
\end{center}
\end{figure}

\begin{figure}[!hp]
\begin{center}
\includegraphics[scale=0.7]{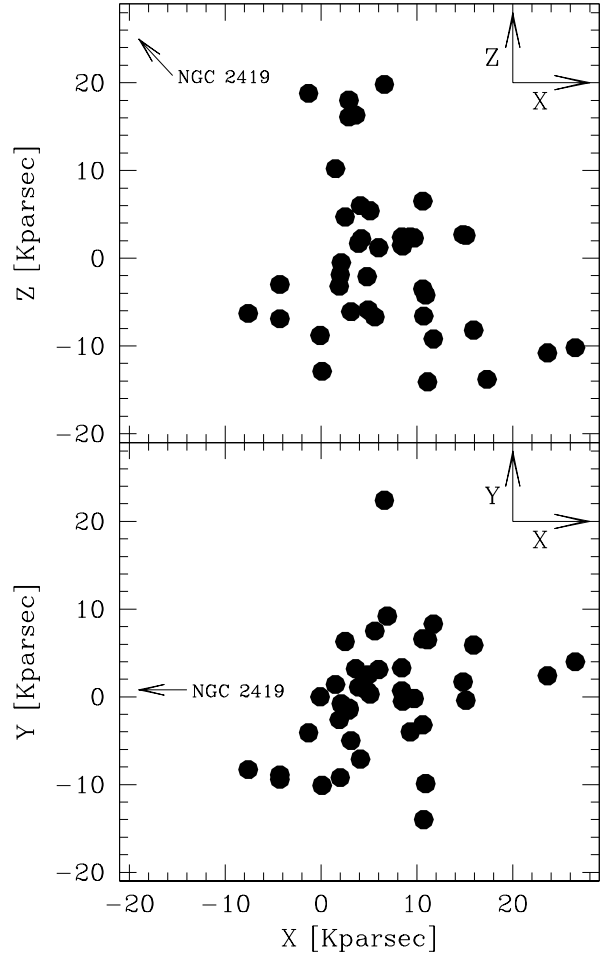}
\caption{Spatial distribution of Galactic GCs in our sample.
Note that NGC~2419, located at a distance of 80\,kpc from the Galactic center, is
outside the plot limits.}
\label{map}
\end{center}
\end{figure}

As mentioned above, our original proposals for Cycles 1 and 4
requested between 1 and 2 orbits to be spent on each cluster, which
would have resulted in maximum exposure times of approximately
3,000\,s on both bands for each target.  However, our program
benefitted from the complexities of GALEX queue scheduling so that
longer exposure times were achieved for some clusters---in some
cases, such as that of NGC~2298, exposure times were an order of
magnitude longer.  Exposure times are also in general longer in the
$NUV$ than in the $FUV$, which is due to the several events of $FUV$
detector shutoffs caused by over-currents in the $FUV$ detector.
Repeated attempts of collecting $FUV$ data led therefore to an
accumulation of $NUV$ exposures.  In total, our program accrued 340
ksec of open shutter time, or the equivalent of $\sim 226$ GALEX
orbits.  A false color picture of the field containing one of our
clusters (NGC~362) is shown in Figure~\ref{colorpic}. The spatial
distribution of our cluster sample is displayed in Figure~\ref{map}.


\subsection{Photometry and calibration}
\label{phot}

\begin{figure}[!hp]
\begin{center}
\includegraphics[scale=0.7]{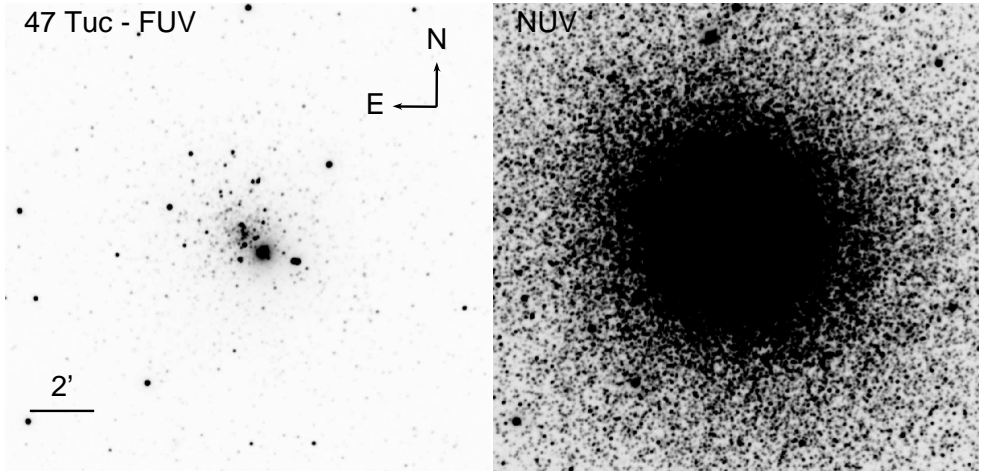}
\caption{$FUV$ and $NUV$ images of 47~Tuc, the reddest cluster in our sample.
Note the vast difference in the crowding of the two images, which
renders $NUV$ photometry impossible in the cluster core, at the
spatial resolution of GALEX.  The $FUV$ light of the cluster is due to a few
dozen sources, with roughly half of it being due to a single very bright
star (47 Tuc BS, O'Connell et al. 1997).  Photometry in the $FUV$ is accurate even in the central cluster
regions.
}
\label{47Tuc_fn}
\end{center}
\end{figure}

\begin{figure}[!hp]
\begin{center}
\includegraphics[scale=0.7]{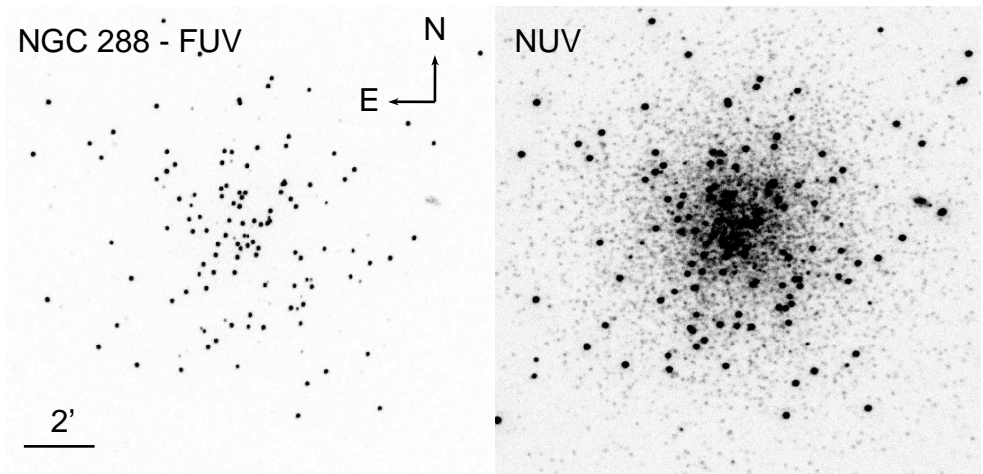}
\caption{$FUV$ and $NUV$ images of NGC~288, with a lower overall surface
brightness than 47~Tuc, yet with a larger population of $FUV$
sources---though not large enough to present problems for $FUV$ photometry in
the central regions.  In the $NUV$, crowding is much more important, yet
photometry at the resolution of GALEX is still achievable, though with
lower precision.}
\label{288_fn}
\end{center}
\end{figure}

\begin{figure}[!hp]
\begin{center}
\includegraphics[scale=0.7]{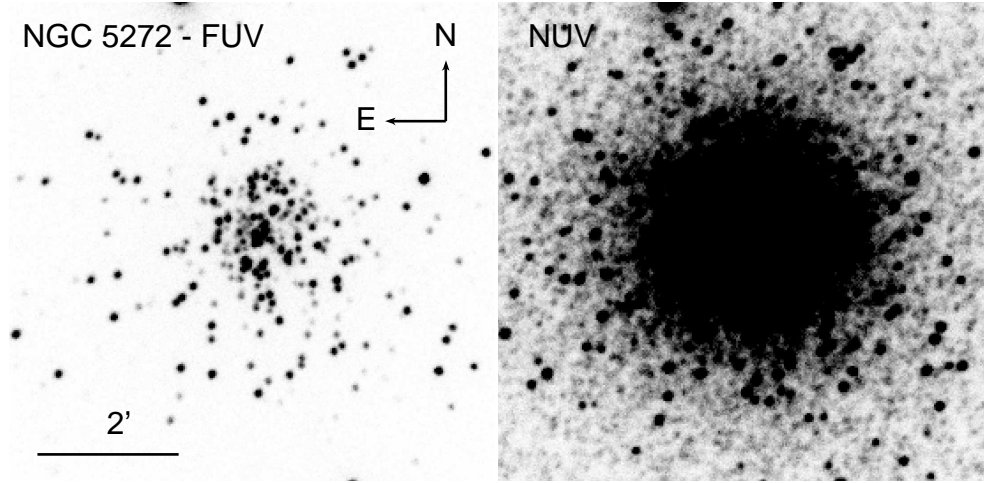}
\caption{$FUV$ and $NUV$ images of NGC~5272. Note that stellar density is high
enough that even $FUV$ photometry is slightly more uncertain than in the
cases of NGC~288 and 47~Tuc.}
\label{5272_fn}
\end{center}
\end{figure}

\begin{figure}[!hp]
\begin{center}
\includegraphics[scale=0.7]{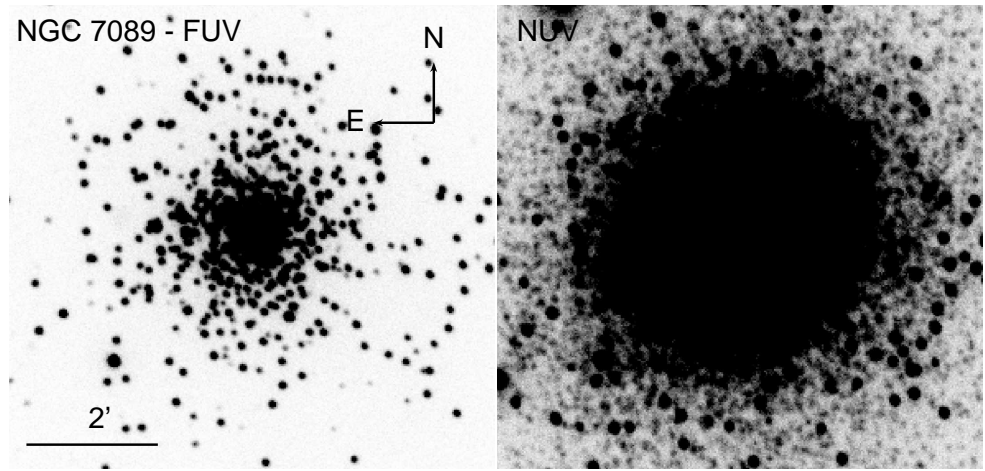}
\caption{$FUV$ and $NUV$ images of NGC~7089, one of the densest clusters in our
sample.  In this extreme case, even $FUV$ photometry is hampered in the core cluster
regions.}
\label{7089_fn}
\end{center}
\end{figure}

The quality and depth of our data are illustrated in Figures~\ref{47Tuc_fn}
to \ref{7089_fn}, where $NUV$ and $FUV$ images of a representative
subset of the sample are displayed.  Consistent gray scales were
adopted when producing these images, to allow for a fair visual
assessment of the various degrees of crowdedness of Galactic GCs,
as seen on both GALEX bands.  To the same end, image sizes are set
such that the field of view is equal to $\sim 5$ times the cluster's
core radii.  Displayed are one of the reddest clusters in our sample
(47~Tuc) together with bluer clusters, spanning a range of stellar
density, increasing from NGC~288 to NGC~5272, and NGC~7089.  As a
general rule, one can see that stellar density is significantly
higher in the $NUV$ than in the $FUV$.  This is because, on one hand,
the combination of higher sensitivity and longer exposure time makes
$NUV$ images a lot deeper than their $FUV$ counterparts and, on the
other hand, Galactic GC stars are predominantly brighter in $NUV$
than in $FUV$.  As a result, even at the relatively low resolution
of GALEX, accurate $FUV$ photometry can be obtained down to the cores
of most clusters in our sample (47~Tuc and NGC~288 being two cases
in point), getting progressively difficult at increasing cluster
density, up to a limit where crowding becomes a problem in the
cluster central regions (e.g., NGC~7089).  Unlike the $FUV$, crowding
in the $NUV$ is a problem in the central regions of almost all clusters
in our sample.  The effect of crowding on our CMDs is discussed in
Section~\ref{cmds}.

The photometric analysis was performed on the background-subtracted
intensity images output by the GALEX pipeline \citep[M07]{mo07}.
These are $3840\times3840\,{\rm pixel}^2$ images with a plate scale of
1.5$\arcsec\,{\rm pixel}^{-1}$, covering a circular area of $~1\fdg2$
in diameter, flat-field corrected and with the flux normalized by
the effective area and exposure time.  Photometry was performed by
following standard procedures for point spread function (PSF)
modeling, using the crowded-field photometry package {\tt DAOPHOTII}
\citep{st87} for both $FUV$ and $NUV$ images.  The first step consists
of defining a number of bright stars across the FOV for PSF modeling.
For that purpose, we performed a very shallow search for bright
point sources with the {\tt DAOPHOT} task {\tt find} on each image.
Magnitudes at this stage, before PSF modeling, were based on simple
aperture photometry obtained using the task {\tt photometry} with
an aperture radius $r=4.5\arcsec$.  We then selected relatively
isolated bright stars spread across the FOV, for PSF determination.
We avoided stars in the very central and crowded regions.  

The PSF model was typically based on 200 stars in $NUV$ and
50 in $FUV$ images.  Quadratic spatial variations of the PSF model
were considered.  Once the PSF model was determined, we reran {\tt
find} and {\tt photometry} with appropriate threshold levels
(typically 3--6~$\sigma$ the sky background) in order to
generate a more complete and deeper list of target stars for photometry.
Magnitudes were then obtained for this expanded list by performing
PSF fits using the {\tt allstar} routine.  Tests were performed
where PSF photometry was carried out replacing the PSFs derived in
this analysis by the average PSFs supplied by the GALEX team\footnote{See
{\tt http://www.galex.caltech.edu/researcher/techdoc-ch5.html}}.
No significant differences between these tests and the original
photometry were found for a couple of clusters bracketing the full
range of stellar densities and number of clean stars available for
PSF determination.

Aperture corrections (typically $\sim 0.2$\,mag) were calculated
on each image by using 15--20 isolated and bright stars, which were
used to generate reliable curves of growth.  Instrumental magnitudes
were converted to the ABMAG photometric system, using the zero points
provided by M07, as follows:

\begin{equation}
FUV = -2.5\,\log ({\rm Counts\,\,s}^{-1}) + 18.82
\end{equation}

\begin{equation}
NUV = -2.5\,\log ({\rm Counts\,\,s}^{-1}) + 20.08
\end{equation}

\begin{figure}[!hp]
\begin{center}
\includegraphics[scale=0.7]{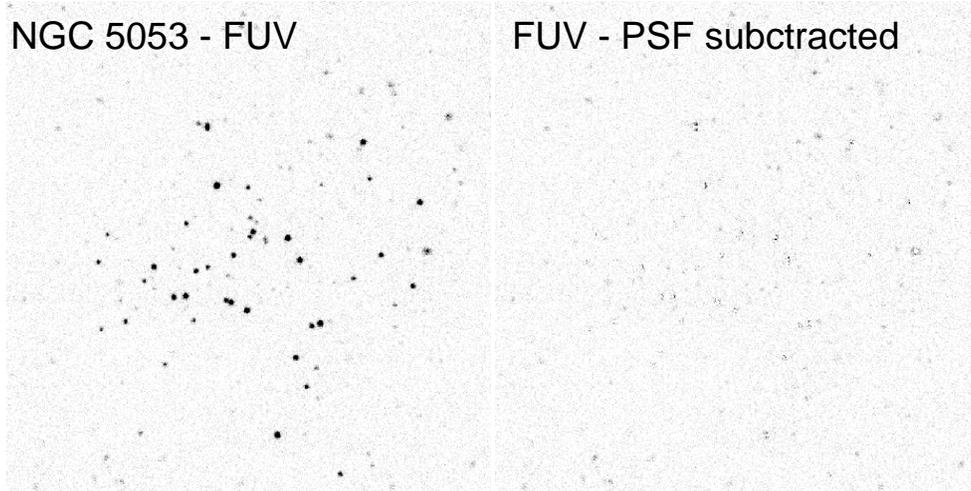}
\caption{Quality of PSF modeling in a good case: NGC~5053 in the
$FUV$.  Due to low density, PSF residuals are negligible and good
quality photometry is achieved for all cluster stars in the $FUV$.}
\label{5053sub}
\end{center}
\end{figure}

\begin{figure}[!hp]
\begin{center}
\includegraphics[scale=0.7]{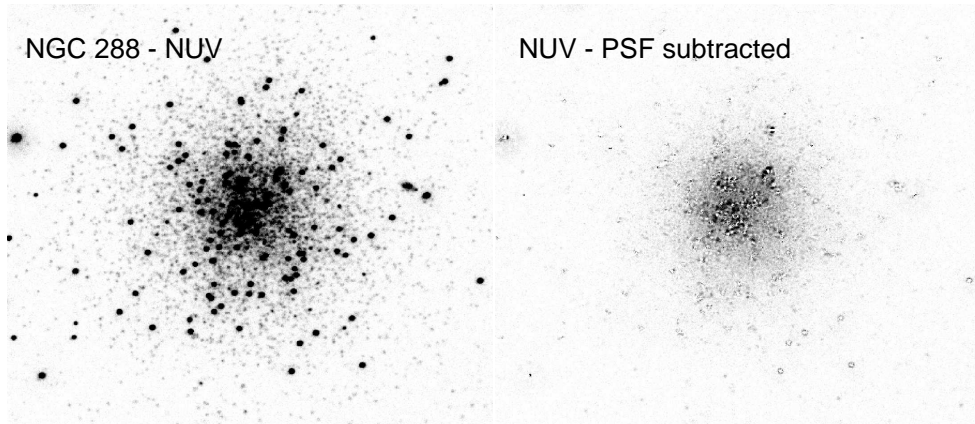}
\caption{Quality of PSF modeling in a typical case: NGC~288 in the NUV.
Crowding in the central regions is relatively high, at the resolution of GALEX,
photometry is relatively inaccurate for stars within $\sim 0.5\arcmin$ from the
cluster center.  The PSF-subtracted image shows a diffuse residual,
associated with the detection of $NUV$ light from unresolved turnoff stars.
}
\label{288sub}
\end{center}
\end{figure}

To illustrate the quality of the PSF modeling, we show in
Figures~\ref{5053sub} and \ref{288sub} typical residuals from PSF
subtraction in $FUV$ and $NUV$ images, respectively.  Visual inspection
shows that in low density areas, such as the $FUV$ image of NGC~5053
in Figure~\ref{5053sub} and the $NUV$ image of NGC~288 outside the
cluster core in Figure~\ref{288sub}, stellar brightness profiles
are properly reproduced by the PSF-models used.  In contrast,
residuals are much worse in crowded areas such as the core of NGC~288
in $NUV$.  For the reasons explained above, at the low spatial
resolution of GALEX, crowding often caused photometry near the
cluster center to be unreliable.  We therefore exclude stars located
within a given cluster-centric distance, for which we felt that
reliable magnitudes could not be obtained on the basis of PSF-fitting
photometry.  The threshold cluster-centric distance varies from cluster
to cluster depending on the density of UV sources.  Moreover, because
crowding was far more severe in the $NUV$ than in the $FUV$, we usually
adopted different cluster-centric distance thresholds for the two
bands.  


Finally, we point out that the outer $\sim 5\arcmin$ annulus of
GALEX images is affected by optical distortions that may cause false
detections and large magnitude errors \citep[see, e.g.][]{re07}.
We note that our photometry is not affected by these problems,
because typically this area is well beyond the tidal radii of the
clusters in our sample, except for the cases of 47~Tuc and M~3
(NGC~104 and 5272, respectively).  For these two clusters, all
photometry within the outer $\sim 5\arcmin$ annulus was discarded.

Cross-correlation of the $FUV$ and $NUV$ catalogs was performed
using CataXcorr, developed at the Observatory of Bologna (P.
Montegriffo et al. 2003, private comunication), which has the
important advantage of allowing a visual check of the quality of
the geometric roto-translation solution.  The final catalogues
consist of stars detected in at least one of the two filters.  This
choice has been made in order to maximize the number of sources for
possible cross-match with optical catalogues.  For the reasons
explained above, there is a large number of $NUV$ sources without
a $FUV$ counterpart.  On the other hand, because crowding is more
severe in the $NUV$ than in the $FUV$, there is a (small) number
of central $FUV$ sources without reliable $NUV$ magnitudes.

\begin{figure}[!hp]
\begin{center}
\includegraphics[scale=0.7]{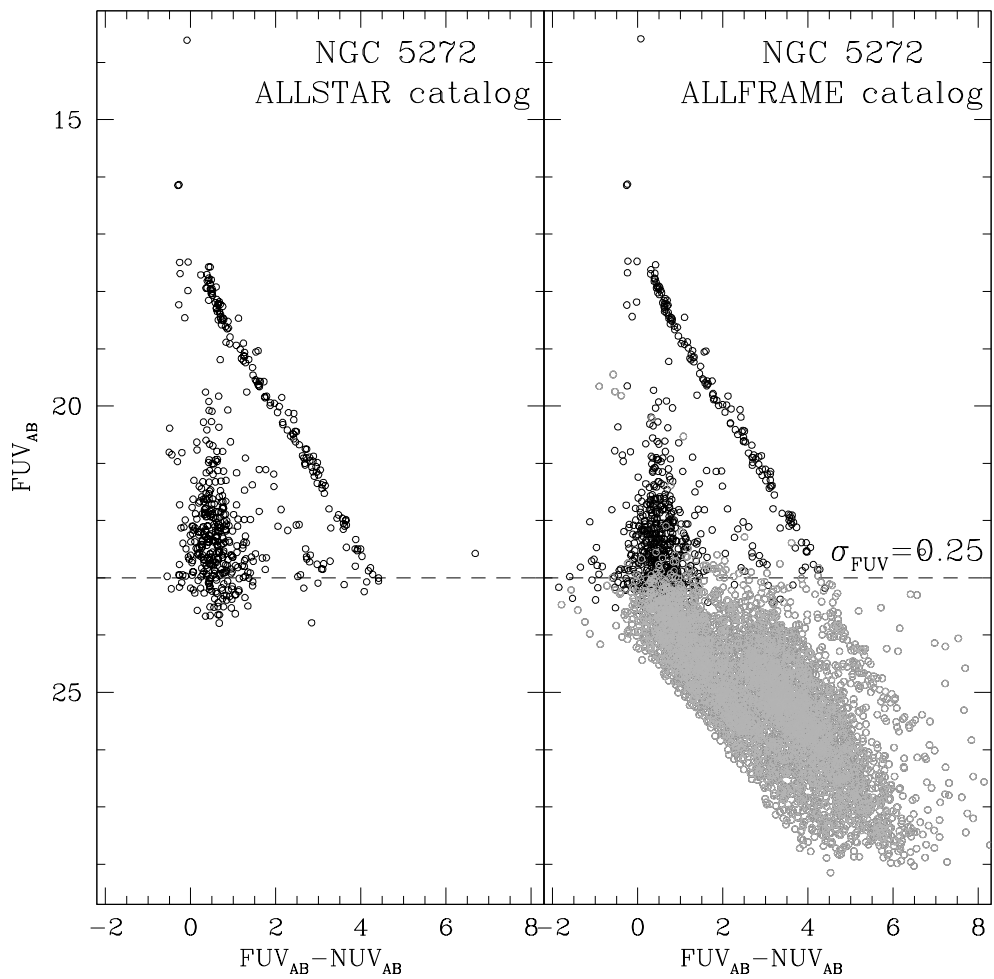}
\caption{Results from tests aimed at assessing the depth of our photometry.
The left panel shows the CMD based on {\tt allstar}, while the right panel 
shows the CMD obtained by ``force-finding'' stars in $FUV$ on the basis of their 
position in $NUV$ by using
{\tt allframe}. The dashed horizontal line marks the limit corresponding to
the photometric error $\sigma_{FUV}=0.25$.} 
\label{m3_comp}
\end{center}
\end{figure}

For the reasons discussed above, the depth achievable in GALEX CMDs
is set by the shallower $FUV$ photometry.  In fact, we showed in
previous works \citep[e.g.,][]{la07,da09} that our $NUV$ images are
often deep enough to detect stars $\sim 1$ mag fainter than the
main-sequence turnoff.  With the aim of maximizing the number of
stars with magnitude measurements in both GALEX bands, we attempted
to use the {\tt allframe} routine (Stetson et al. 1989) in order
to ``force-find'' stars in the $FUV$ images on the basis of their
positions in the $NUV$.  Results of this experiment are shown in
Figure~\ref{m3_comp}.  The left panel shows the CMD from photometry
based on {\tt allstar}, and on the right panel the CMD obtained
from forcing {\tt allframe} to find $FUV$ stars at their known $NUV$
positions is shown.  These plots suggest that this use of {\tt
allframe} leads to detection of sources ~3--4 mag fainter in $FUV$
than just using {\tt allstar}.  However by performing a visual
analysis on the images, it became clear that most of the additional
$FUV$ detections were not real.  To further verify this result, we
performed PSF photometry at random $FUV$ background positions ending
up with a color-magnitude distibution that is very similar to the
one obtained when force-finding $FUV$ stars (gray points in the
right panel of Figure~\ref{m3_comp}).  For this reason we decided
to adopt only the catalogues obtained by using the {\tt allstar}
routine as already described.  Fig.~\ref{m3_comp} shows also that
the two approaches give virtually identical results when stars with
$\sigma_{FUV}>0.25$ are removed from the CMD.

Photometric depth varies from cluster to cluster according to
exposure times (see Table~\ref{targets}), thus being in all cases
deeper in the $NUV$ than in the $FUV$.  In our deepest images, we
reach $FUV \sim 24.4$ and $NUV \sim 25.0$.  NGC~2419 is the only
cluster in our sample (with both $FUV$ and $NUV$ images available)
for which it has not been possible to obtain reliable photometry
of individual stars.  Since NGC~2419 is a cluster with a large
population of emitters both at $NUV$ and $FUV$ wavelenghts and it
is one of the most distant clusters in the Galaxy ($d=87\,{\rm
kpc}$, Dalessandro et al.  2008, see Fig.~1), it appears extremely
dense in GALEX images making photometric measurements of individual
stars virtually impossible at the GALEX spatial resolution.

\subsection{Deviations from Linearity} \label{linear}

The GALEX detectors present deviations from linearity when count
rates exceed $\sim 1000\,{\rm counts\,s}^{-1}$ (see M07). This
affects bright source photometry, particularly in the $FUV$ (see
below).  In order to correct observed magnitudes, M07 compared
aperture photometry for a sample of Hubble Space Telescope (HST)
spectrophotometric standards observed by GALEX, with synthetic
photometry based on spectrophotometric data from the CALSPEC
database\footnote{See
http://www.stsci.edu.instruments/observatory/cds/calspec.html}.
Because our photometry is based on PSF-fitting instead of aperture
photometry, we decided to repeat the analysis done by M07, by
performing PSF photometry on the GALEX archival data for HST
spectrophotometric standards, in order to assess the impact of
deviations from linearity on our magnitudes.  We used 13 of the
spectrophometric standards from M07 (see Table~\ref{standards})
spanning a range of 4--5\,mag both in $FUV$ and $NUV$. For each of
these stars we obtained $FUV$ and $NUV$ magnitudes by using the
same procedures described in Section \ref{phot}, and compared our
results with those from M07.

\begin{figure}[!hp]
\begin{center}
\includegraphics[scale=0.7]{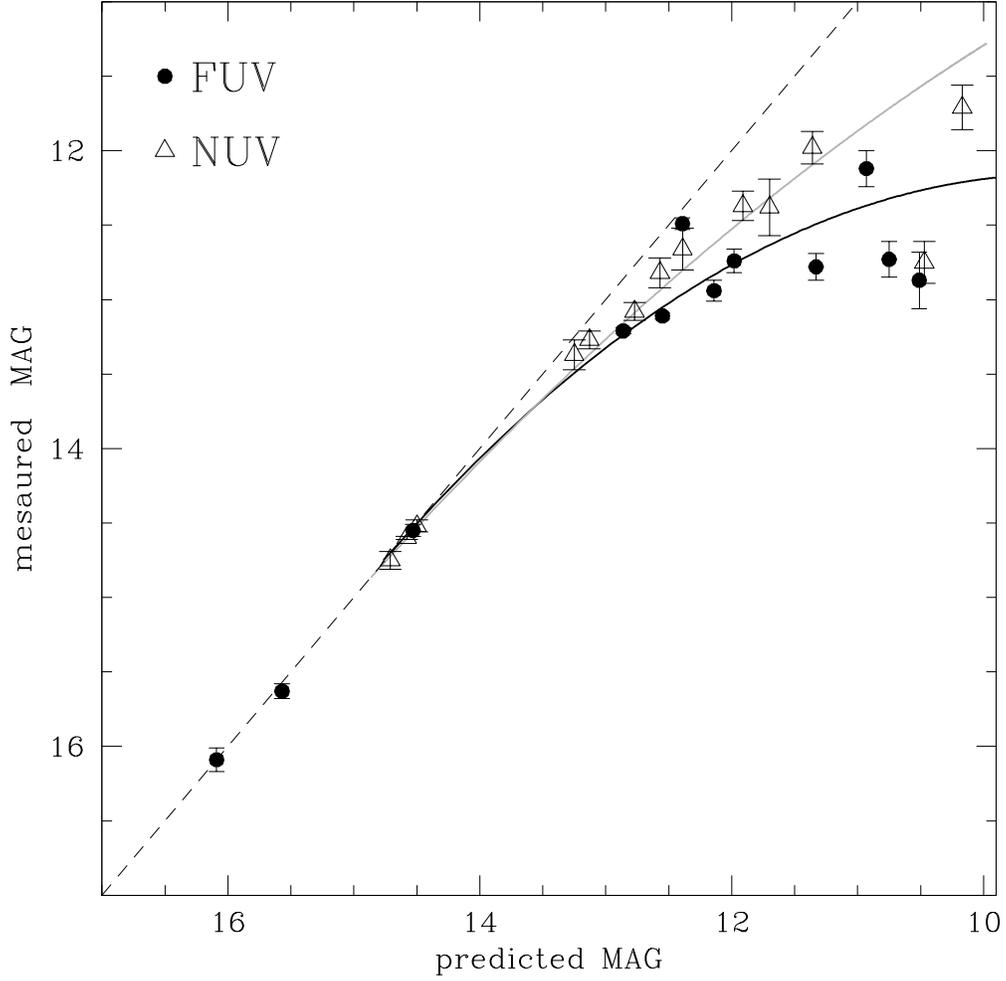}
\caption{Results of tests for deviations from linearity in GALEX detectors.
A comparison is shown between ``predicted'' and measured magnitudes for a
number of standard stars from the CALSPEC database (see text).
``Predicted'' values are magnitudes obtained through synthetic photometry
on CALSPEC spectra by M07, while measured values are aperture magnitudes by
M07 (curves) and PSF magnitudes from this work (data points).  Gray (black)
curves and open (filled) symbols represent $NUV$ ($FUV$) magnitudes.
Saturation becomes important at 14.5\,mag for both channels alike, but is
more intense in the $FUV$ channel for brighter sources.  Our PSF photometry
seems to be only slightly more affected than M07's aperture photometry, in
the $FUV$ channel only.  Only a handful of stars in our entire sample are
substantially affected by detector non-linearity.
}
\label{lin1}
\end{center}
\end{figure}

The results are displayed in Figure~\ref{lin1}, where our measurements
are plotted against synthetic magnitudes as reported by M07.  Data
points for both $FUV$ (filled circles) and $NUV$ (open triangles) are
plotted.  The solid lines are fits from M07 to the relation between
their aperture magnitudes and synthetic photometry, the black
(gray) line represents fits to $FUV$ ($NUV$) data.  The dashed line
shows the one-to-one relation.  It is clear from this figure that
non-linearity becomes detectable in both bands at $\sim 13^{th}$\,mag.
Deviations increase with increasing brightness, the effect being more
severe in the $FUV$ than in the $NUV$.  For the brightest $FUV$
source, non-linearity leads to a 2.5\,mag overestimate in magnitudes,
the effect being $\sim 1$\,mag weaker in the $NUV$.  Most importantly,
all but a handfull of the stars for which we have photometry are
safely below the limit where non-linearity effects are detectable.

It is interesting to contrast our results with those by M07, by
comparing our data points with their fits in Figure~\ref{lin1}.  In
the $NUV$ case, deviations from linearity are consistent between this
work and M07, including a star that deviates very strongly from
linearity (BD33, in Table~\ref{candidates}), for which our photometry is
in good agreement with that of M07.  On the other hand, the
data suggest that non-linearity effects are slightly stronger in
our PSF-fitting photometry than in M07's aperture photometry,
particularly in the $FUV$.

\section{The Color-Magnitude Diagrams} \label{cmds}

The CMDs obtained in this work are displayed in Figures~\ref{to1}a-g.
The outstanding variety of colors and magnitudes of UV bright sources
in Galactic GCs is immediately obvious, even on a perfunctory perusal
of these diagrams.  There are, nonetheless, features that are common
to all diagrams, and we briefly comment on those here.  In PAPER
III, we present a new classification of Galactic GCs, based on the
morphology of their HBs in UV CMDs, and study correlations between
this new HB morphology index with global cluster properties.

We start by discussing the CMD of M~3 (NGC~5272), which is reproduced
in better detail in Figure~\ref{5272}.  M~3 is a moderately metal-poor
Galactic GC (${\rm [Fe/H]} \sim -1.5$) with a relatively blue HB
\citep[HB parameter = 0.08,][]{le94,bm00}.  GALEX magnitudes were
corrected from extinction values estimated using the \cite{ca89}
extinction curve and extinction values from \cite{ha96}.  Extinction
in the UV is substantially higher than in the optical, amounting
in the case of NGC~5272 to $A_{FUV} \sim 0.08$ and $A_{NUV} \sim
0.09$, as opposed to $A_V \sim 0.03$.  We note however, that, because
the effective wavelength of the $NUV$ filter coincides with a bump
in the Galactic extinction curve \citep{ca89}, interstellar extinction
does not redden $FUV-NUV$, leading instead to a slight blueing of
that color.  Finally, absolute magnitudes in Figure~\ref{5272}
were obtained adopting distance moduli taken from \cite{ha96}.

Only stars located at cluster-centric distances between $120\arcsec$
and $1300\arcsec$ are displayed in Figure~\ref{5272}, to minimize
crowding effects on photometry performed within the cluster core,
and to minimize field contamination beyond the cluster tidal radius.
A $T_{\rm eff}$ scale is provided on the top axis of the diagram,
which was obtained by interpolating values into {($FUV - NUV$)} vs.
$T_{\rm eff}$ vs.  [M/H] tables calculated on the basis of fluxes
from Kurucz model atmospheres,\footnote{see
http://kurucz.harvard.edu/grids.html} adopting the filter responses
available on the GALEX website.  Models were adopted for surface
gravities typical of HB stars \citep{do93}, so that the scale does
not apply in detail to other stellar types such as blue stragglers
and PAGB stars.

The first vacuum-UV CMDs for globular clusters were obtained by the
{\it Astro}/Ultraviolet Imaging Telescope \cite[e.g.,][]{hi92,pa94,wh94},
and the general features of those diagrams are also seen in our
GALEX photometry.  Some of those same features are also seen in HST
CMDs obtained by \cite{fe97} and \cite{fe03}.  Three main structures
are visible in the CMD of this cluster, as indicated in Figure~\ref{5272}.
The cluster HB extends from the lower right to the upper left of
the diagram, ranging from 4.5 to $-0.5$ in $(FUV-NUV)$, and from 8
to 2 in $M_{FUV}$.  It is obvious from this figure that the
``horizontal'' branch is not horizontal in the UV (slightly more
so in $NUV$ than $FUV$), and its slope is mainly a result of
bolometric correction effects.  The HB spans a wide range in $T_{\rm
eff}$, going from F stars in the blue HB, at $T_{\rm eff} \sim
7,000$\,K, all the way to O stars in the so-called extreme HB at
$T_{\rm eff} \sim 30,000$\,K.  A few stars are also seen at the
blue end of the HB, displaced by up to 1 mag fainter in FUV than the
blue tip of the horizontal branch, at about $M_{FUV} \sim 3$ and
$FUV_{AB}-NUV_{AB} \sim -0.25$.  Those are the so-called ``Blue
Hook'' stars, whose origin is still not well understood
\citep[e.g.,][]{wh94,mo04,bu07,ro08}.

A few gaps are apparent along the HB of Figure~\ref{5272},
one of them at $(FUV-NUV)/T_{\rm eff} \sim 0.9/8,500\,{\rm K}$, and
two other less prominent ones located at $(FUV-NUV)/T_{\rm eff}
\sim 3.3/7,450\,{\rm K}$ and 0.0/12,000~K.  The latter gap is the
one that is the most likely to be real.  It corresponds to the ``G1''
gap, identified by \cite{fe98} in HST/WFPC-2 optical color-magnitude
diagrams of M~3 and other Galactic globular clusters.  It also
coincides with the position associated with the Grundahl jump---a
discontinuity in the HBs of globular clusters, first
pointed out by \citep{gru98}, which manifests itself as a brightening
of the Str\"omtren $u$ or the Johnson U band magnitudes of stars hotter
than $T_{\rm eff} \sim $ 11,500~K.  The Grundahl jump has been
interpreted by \cite{gru99} as being due to a decrease of hydrogen-
relative to metal-opacity, associated with an increase of light
element opacities due to radiative levitation for $T_{\rm eff}
\simgreater $ 11,500~K.  Inspection of Figure~8 of \cite{gru99}
suggests that the differential impact of radiative levitation on
FUV and NUV-like photometric bands can potentially generate a gap
with a similar size to that observed in Figure~\ref{5272}.  However,
a definitive association between this apparent gap and the Grundahl
jump depends on currently unavailable synthetic photometry based
on detailed model atmosphere calculations for the relevant stellar
parameters and abundance patterns.

The remaining two gaps do not seem to have observed counterparts
in the CMDs of \cite{fe98}, which casts doubts on the reality of
those gaps.  As pointed out by \cite{ca08}, stochastic effects due
to small samples could be to blame, since some of the previously
proposed gaps did not stand the test of better quality color-magnitude
diagrams, based on more robust samples.  According to \cite{ca08},
real features such as the Grundahl jump are probably associated to
chemical composition discontinuities along the HB, which can manifest
themselves through opacity effects due to specific chemical species,
which may operate on some photometric bands, but not on others.
The latter could conceivably explain the presence of these two gaps
in our CMD, but not in those of \cite{fe98}, provided an opacity
source can be identified that is important in the NUV/FUV but not
in the optical.  Alternatively, these gaps may be due to the fact
that the nonlinearity of the $(FUV-NUV)$--$T_{\rm eff}$ relation
leads to a color stretching of the redder part of the UV HB, which
may make such gaps more readily detectable in the UV than in the
optical.  This issue clearly deserves further investigation in
future studies.

Another feature of UV HB morphologies is the clump of stars at $2.5
\simless M_{FUV} \simless 3.5$ and $0.3 \simless {FUV}-{NUV} \simless
0.8$.  This feature is actually an artifact caused by the highly
non-linear character of the color-${T_{\rm eff}}$ relation.  At
$({FUV}-{NUV}) \sim 4.0$ ($T_{\rm eff} \sim  7,000$\,K), a 0.5\,mag
color interval spans a few 100\,K in $T_{\rm eff}$, whereas at
$({FUV}-{NUV}) \sim 0.3$ ($T_{\rm eff} \sim 10,000$\,K) the same
color interval spans several 1,000~K, leading to the accumulation
of data points in that area of the HB for any cluster with a
substantial number of stars hotter than $T_{\rm eff} \sim 8,500$\,K.

The next important population visible in the CMD of Figure~\ref{5272}
is that of blue stragglers.  Their identification in this case is easy,
as they are spread along a sequence that is parallel, and 1--1.5
mag fainter than the HB (e.g. Ferraro et al. 1999; see also Figure 2 by Ferraro
et al. 1997).  A \cite{gi00} zero-age
main sequence for the metallicity of NGC~5272 is plotted as a dashed
line, in order to facilitate the identification of the cluster's
blue stragglers.  Only the hottest and brightest blue stragglers
are detected in the $FUV$.  \cite{la07} and \cite{da09} have recently shown
that combination of GALEX data with wide-field optical photometry 
is a powerful mean to study blue stragglers, and in particular their
spatial distribution in GCs.

Another important population in this CMD is that of post He-core
burning stars, whose identification is difficult, given their rarity
and the uncertainties surrounding their evolutionary paths in the CMD,
as well as their lifetimes.  There are two PAGB candidates in this
CMD, which are approximately 1.5\,mag brighter than the brightest
HB stars, at $T_{\rm eff}$ greater than $\sim 20,000$\,K.  See
discussion in Section~\ref{uvbright}.

The cloud of points that is located towards fainter magnitudes and
bluer colors than the HB is mostly populated by background sources,
with an average color of $(FUV-NUV) \sim 0.5$ and $M_{FUV} \simgreater
5$ ($FUV \simgreater 19$ in Figures~\ref{to1}a-g).  Some
of those objects may actually belong to the cluster populations
with bright blue stragglers contributing on the red side and young
white dwarfs demarcating the blue envelope.  Based on WFPC2 data,
\cite{fe01} argued for the presence of young white dwarfs, with
ages $\simless 13$ million years, in the corresponding locus of the
$(m_{F218W}-m_{F439W})$ CMD of 47~Tuc.  In particular, they showed
that the blue envelope of that CMD population is consistent with
theoretical expectations both for the colors and number counts of
young white dwarfs.  However, while that study refers to a small
region at the center of the cluster, where the background field
contamination is expected to be low, the GALEX FoV is expected to
be heavily contaminated by background objects.  In fact, inspection
of high resolution images taken with the wide field imager, attached
to the ESO 2.2~m telescope \citep{la07}, indicated that the majority
of the sources in that region of the CMD consists of distant galaxies.
In addition, the number of objects in this region of the UV CMD of
M~3 is consistent with the number of extra-galactic known objects
as found in the NASA Extragalactic Database
(NED)\footnote{http://nedwww.ipac.caltech.edu/}. In summary, the
low resolution of GALEX images and the relatively low resolution
of the ESO 2.2\,m images do not allow one to distinguish unequivocally
between white dwarfs and blue stragglers on one side, and background
galaxies on the other.  Therefore, we decide to leave them in the
plots, with the caveat that absolute magnitudes should be disregarded
for most objects in that region of this diagram.

The effect of crowding
on the GALEX CMDs can be assessed in Figure~\ref{7089}, where stars
in the field of NGC~7089 are plotted.  Stars within 2\arcmin\ from the
cluster center are shown as gray triangles, whereas stars at larger
cluster-center distances, within the cluster tidal radius, are
plotted with open circles.  The different CMD loci occupied by stars
within and outside the 2\arcmin\ radius shows that crowding produces a
population of stars artificially brighter and redder than the
cluster's HB population.  While the brighter magnitudes
are a straightforward effect of blending, the apparent redder colors
are due to the fact that blending is more severe in the $NUV$ than
in the $FUV$.  The case of NGC~7089 is somewhat extreme, since the
HB of this cluster is so populous that crowding is
important in both $FUV$ and $NUV$ images (Figure~\ref{7089_fn}).  In
most cases, crowding in the $FUV$ is far less severe, and its effect
on CMDs is that of producing a predominantly redder population, due
to crowding in the $NUV$.  We also point out that because the
``brightening'' effect associated with stellar blending should be
typically of the order of 0.75\,mag, it is possible that some of the
very bright stars at $FUV \simless 14.5$ in NGC~7089 may be
real UV-bright cluster members.  See discussion in Section~\ref{uvbright}.


\section{UV-Bright stars}  \label{uvbright}

While the integrated light of old stellar populations in the $FUV$
is dominated by EHB stars, post-He core burning stars also contribute
a fraction of that radiation \citep[e.g.,][]{gr90,oc99,gr99}.  
A few definitions are required at this point.  According to standard
stellar evolution theory, post-HB evolution depends strongly on the
mass of the stellar envelope.  After core-He exhaustion, stars with
the highest envelope masses evolve into the AGB phase, undergoing
thermal pulses and eventually losing their envelopes, evolving
towards higher temperatures at constant high luminosity as PAGB
stars.  Stars with lower envelope mass experience a much shier
excursion into the AGB phase and never undergo thermal pulses,
evolving towards higher temperatures, after envelope loss, at
constant, but lower, luminosities.  The latter are called post
early-AGB (PEAGB) stars.  Finally, at the extreme low end of envelope
mass, stars never make it to AGB phase after core-He exhaustion,
departing the blue end of the HB in a small excursion towards higher
luminosities, but never becoming as bright as PEAGB stars.  The
latter are the so-called AGB-manqu\'e stars (AGBM).

Our knowledge of the total contribution of these stars to the
integrated light of old stellar populations is limited by uncertainties
in evolutionary tracks, which are to a large extent due to difficulties
in the modeling of mass loss during the AGB phase \citep{vw03}.
GCs are the one type of stellar systems where the
masses of these stars are best constrained, so that observations
of post core-He burning stars in clusters can in principle contribute
to the betterment of stellar evolution models.  However, stellar
evolution proceeds at a very fast pace after the core-He burning
stage, with time scales varying between $10^4$ and $10^6$ yr.  The
incidence of these stars in stellar systems of relatively low mass,
such as GCs, is therefore low, and thus strongly
affected by stochastic effects.  The wide field of view of GALEX
and the size of our sample configure an ideal situation for the
cataloguing of these rare stellar types.  We describe in this Section
the procedure we followed in order to idenfity PAGB and other
UV-bright star candidates.

\begin{figure}[!hp]
\begin{center}
\includegraphics[scale=0.9]{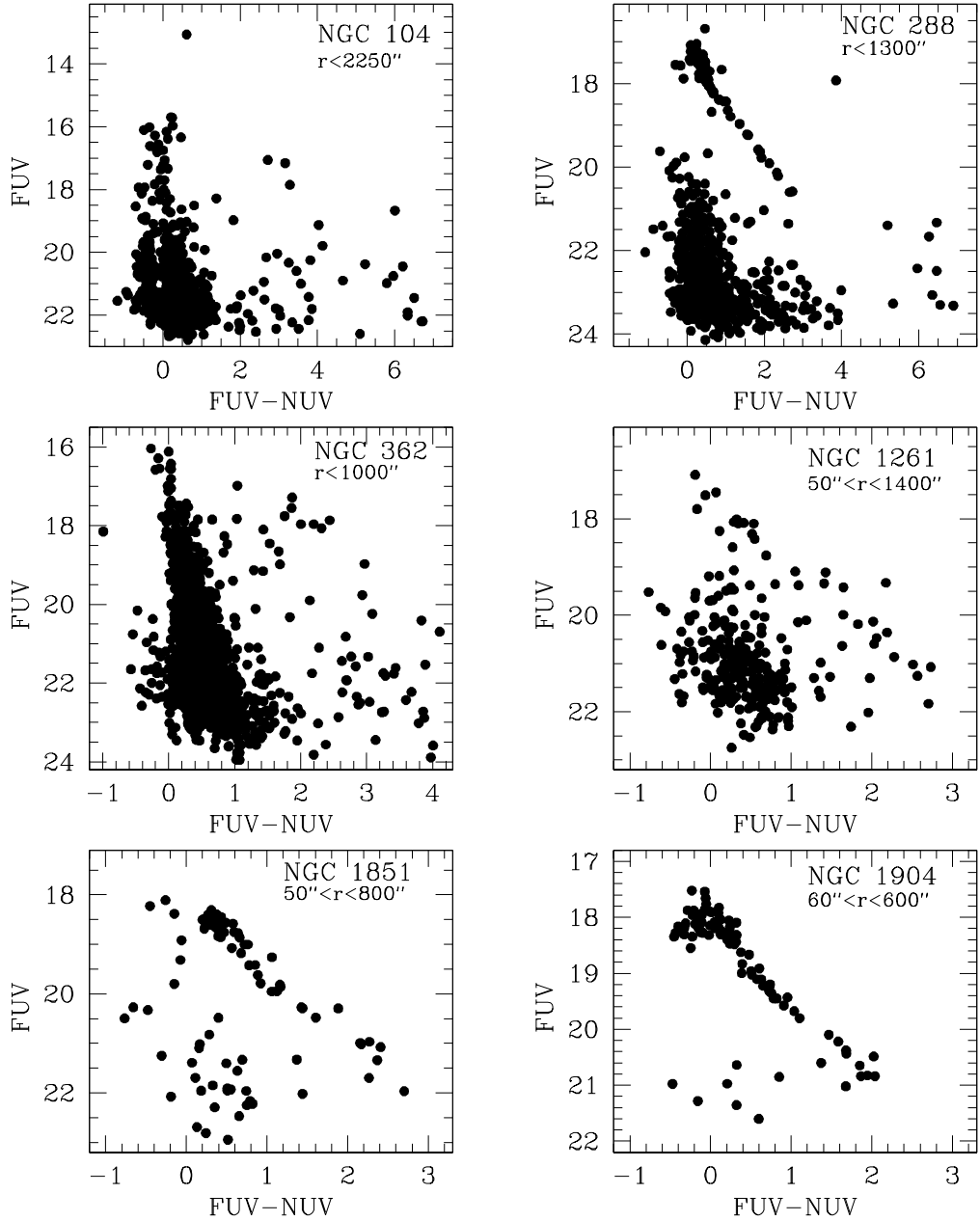}
\caption{a. Color-magnitude diagrams of Galactic GCs.  The photometry
shown in this Figure through Figure~\ref{to7} was {\it not} corrected
for reddening or extinction.  Note the variety of HB morphologies.
The vast majority of the objects in the CMD of 47~Tuc (NGC~104) and
NGC~362 with $(FUV_{AB} - NUV_{AB}) \simless 1.5$ are actually
main-sequence stars from the Small Magellanic Cloud.  
}
\label{to1}
\end{center}
\end{figure}

The paucity of post-core-He burning stars makes their identification
solely on the basis of photometry in any given single GC extremely
uncertain, though an early attempt was made using a UV CMD of NGC~6752 by
\cite{la96}.  Because the average number of PAGB stars per cluster
is of the order of $\sim$ 1, they form no {\it sequence} in any of
the CMDs shown in the previous section.  In the absence of a sequence,
distinguishing post-core He burning stars from fore/background field
contaminants in the CMD of any individual cluster is very hard, and
usually requires a spectroscopic follow up.  However, stacking the
CMDs of many clusters should boost the number of UV-bright stars
per unit CMD area, highlighting the locus occupied by stars in these
evolutionary stages.  Figure~\ref{pagb_data} shows a stack of the
best 23 CMDs from Figures~\ref{to1}a-g, which do not have
a very strong background contamination.  The clusters included are
NGC~1851, 1904, 2298, 4147, 4590, 5024, 5053, 5272, 5466, 5897,
5904, 6101, 6218, 6229, 6254, 6341, 6535, 6584, 6809, 6981, 7089,
7099, and 7492.  Data for each cluster were placed on an absolute
magnitude scale, adopting reddening and distance modulus from the
latest version of the \cite{ha96} catalog, and only stars within
the radial limits displayed in Figures~\ref{to1}a-g are
shown in Figure~\ref{pagb_data}.  The spread in magnitude of the
stacked HB is likely caused by uncertainties in the adopted distance
moduli and in the adopted reddening values.

Because all the clusters are brought to the same distance, all the
typical features of the UV CMDs of old stellar populations appear
in sharp contrast in this CMD stack.  For instance, the blue straggler
sequence stretching below the red part of the HB, and the supra-HB
stars in the other extreme of the HB are more clearly seen in the
CMD stack than in most individual CMDs of Figures~\ref{to1}a-g.
Two stellar sequences brighter than the HB are also
apparent in Figure~\ref{pagb_data}.  The bluest and brightest in
$FUV$ have colors roughly between $(FUV-NUV) = -0.5$ and $+0.5$,
and extend to magnitudes as bright as $M_{FUV} \sim -3$.  The other
family of stars is located towards redder colors and fainter
magnitudes, consisting of a population of stars on average 2--3 mag
brighter than the HB, with $(FUV-NUV)$ $\simgreater$~1.5.  These
stars are mostly foreground contaminants, as discussed below.
Finally, we note that there is a population of stars that are
brighter than the HB by no more than 1\,mag, spread through its
entire extension.  These are most likely unresolved stellar blends.

We first turn our attention to the main objects of interest, the
population of stars revealed by the CMD stack just above the extreme
HB stars at $T_{\rm eff}$ of a few times $10^4$\,K.  This stellar
sequence is too blue and extends towards too bright magnitudes to
harbor a significant fraction of blends. We note that in Figure~\ref{7089}
almost all the stars considered to be due to blends produced by
crowding effects are redder than $(FUV-NUV) \sim 0$ (the effect of
interstellar extinction on colors in the CMD of NGC~7089 is
negligible).  Moreover, because Figure~\ref{pagb_data} excludes
stars within central cluster regions, crowding effects should be
minimal anyway.  So, we conclude that this sequence of hot UV-bright
stars constitutes a real population of UV-bright stars hosted by
our sample of Galactic GCs.  In fact, these stars indeed occupy the
same locus as the PAGB, PEAGB, and AGBM stars identified by \cite{br08}
in a STIS UV CMD of M~32 stars (their Figure 3).  In order to gain
further insight into their nature, we reproduce the CMD stack in
Figure~\ref{pagb_mod}, overlaying evolutionary tracks by \cite{br08}
for a PEAGB and a PAGB star of $\sim 0.5$ (dash-dotted line) and
$0.8\,M_\odot$ (thick solid line), respectively.  The model prediction
for the zero-age horizontal branch (ZAHB; dashed line) is also
shown, which matches very well the lower envelope of our observed
HB.

On the basis of the discussion above, we can use the evolutionary
tracks in Figure~\ref{pagb_mod} to assign the UV-bright stars in
our sample to the above evolutionary classes.  Candidates for the
different classes are listed in Table~\ref{candidates} and shown
in Figure~\ref{pagb_mod}, where the data from Figure~\ref{pagb_data}
are shown as gray dots.  Filled circles indicate the positions of
all PAGB candidates in our entire cluster sample, regardless of
their cluster-centric distances.  The large gray triangles indicate
the positions of a few PAGB stars known to exist in clusters from
our sample.  We chose not to impose a cluster-centric distance cut
in our selection of PAGB and PEAGB candidates, because they are
bright enough that crowding effects on their photometry are minimal.
That is not the case of AGBM stars, though, which lie close enough
in magnitude to the HB that their locus in the CMD may be substantially
contaminated by unresolved pairs of HB stars.  Therefore, the list
of AGBM candidates presented in Table~\ref{candidates} only includes
stars within the cluster-centric distance thresholds displayed in
Figures~\ref{to1}a-g.  We consider stars brighter than
the ZAHB by more than 1\,mag in $M_{FUV}$ and fainter than the PEAGB
tracks to be AGB-manqu\'e candidates.  Stars brighter than the PEAGB
class are either PEAGB or PAGB candidates, we therefore refer to
these stars as P(E)AGB.  In view of the uncertainties in evolutionary
tracks and the possible contamination of our magnitudes by stellar
blends (for stars within the crowded areas of the clusters), we
refrain from attempting a distinction between the latter two classes
in our sample.  Finally, stars brighter than the PAGB track are
considered to be PAGB candidates.  Note that two of the stars
identified as PAGB in previous literature (large gray triangles)
would be classified as AGBM and P(E)AGB according to our classification
scheme.  We also impose a color cut in our definition of PAGB,
PEAGB, and AGBM candidates, by requiring that they have $(FUV-NUV)
< 0.7$.  Note that the AGBM and P(E)AGB candidates identified in
Figure~\ref{pagb_mod} include stars from {\it all} clusters in our
sample, not only the 23 clusters included in the stacked CMD from
Figure~\ref{pagb_data}.

Finally, we focus on the redder population of stars brighter than
the horizontal branch.  According to \cite{br08} tracks , PAGB stars
spend only 25\% of their time with colors redder than $FUV_{\rm
AB}-NUV_{\rm AB} \sim 0.7$, so the fact that there are more bright
stars in Figure~\ref{pagb_data} on the red side of that color
threshold than in the blue side is strongly suggestive of the
presence of back/foreground contamination.  There are approximately
26 stars in Figure~\ref{pagb_data} with $M_{FUV} > 2.2$ and $FUV_{\rm
AB}-NUV_{\rm AB} < 0.7$.  Conversely, there are approximately 67
stars with brighter than the HB by $\sim$ 1 mag and with $0.7\
<FUV_{\rm AB}-NUV_{\rm AB} < 5$.  If the evolutionary tracks are
correct, we would expect to find no more than $\sim 9$ stars in
that region of the CMD.  Therefore, we suggest that the vast majority
of the bright stars redder than $FUV_{\rm AB}-NUV_{\rm AB} \sim
0.7$ are not cluster members, likely being foreground A and F stars.
That is not to say, of course, that there are no cluster PAGB stars
in that region of the diagram---in fact, they are very likely to
be there, but finding them on the basis of GALEX data alone would
be like finding needles in a haystack.  Therefore we impose a color
cut in our definition of PAGB, PEAGB, and AGB-manqu\'e candidates,
by requiring that they have $FUV_{\rm AB}-NUV_{\rm AB} < 0.7$.  This
color cut is aimed at minimizing contamination of the candidate
sample by back/foreground contaminants sources.

\section{Conclusions} \label{epilogue}

We have used GALEX to image 44 Galactic GCs in the $FUV$ and $NUV$,
thus creating the largest homogeneous database of the UV properties
of these systems. In this paper we describe the sample selection,
observations, and data reduction, presenting a brief description
of the main features of the UV CMDs.  HB stars are the most important
feature of the UV color magnitude diagrams, and our CMDs reveal an
outstanding variety in the shape of the HB in our cluster sample.
Blue straggler stars are also detected in many clusters.  We present
a catalog of PAGB, PEAGB, and AGBM  candidates, which should be
useful for studies of these rare, but UV-bright, stellar types.  We
hope these data will provide better constraints on models of stellar
evolution during, and after, the HB phase.  In Paper II, we present
the integrated UV photometry for this sample, while a new classification
scheme of the morphology of the HBs of Galactic GCs in UV is presented
in Paper III.

GALEX provided us with an opportunity, unique in this decade, to
collect precious data that will be crucial to help untangling the
intricacies of the latest stages of evolution of low-mass stars,
so as to allow a deeper understanding of the UV properties of old
stellar populations.  We hope that this data set will enable notable
progress in this field during the upcoming years.  The photometric
catalogs can be downloaded from {\tt http://www.cosmic-lab.eu}.


\acknowledgements

Based on observations made with the NASA Galaxy Evolution Explorer.
GALEX is operated for NASA by the California Institute of Technology
under NASA contract NAS5-98034.  We thank Allen Sweigart and Tom
Brown for making available the theoretical evolutionary tracks
employed in this paper.  R.P.S. acknowledges funding by GALEX grants
\# NNG05GE50G and NNX08AW42G and support from Gemini Observatory,
which is operated by the Association of Universities for Research
in Astronomy, Inc., on behalf of the international Gemini partnership
of Argentina, Australia, Brazil, Canada, Chile, the United Kingdom,
and the United States of America.  E.D. thanks the hospitality and
support from Gemini Observatory where some of this work was developed,
during two extended visits in 2009 and 2010, and a shorter visit
in 2011.  This research is also part of the project COSMIC-LAB
funded by the European Research Council (under contract
ERC-2010-AdG-267675).  The financial contribution of the Italian
Istituto Nazionale di Astrofisica (INAF, under contract PRIN-INAF
2008) and the Agenzia Spaziale Italiana (under contract ASI/INAF/I/009/10)
is also acknowledged.  S.T.S. and J.R. are partially supported by
GALEX grant GI3-0075/NNX07AP07G.  S.-C.R. acknowledges support from
Basic Science Research Program (No. 2009-0070263) and th Center for
Galaxy Evolution Research through the National Research Foundation
of Korea (NRF).  S.-J.Y.  acknowledges support from Mid-career
Researcher Program (No. 2009-0080851) and Basic Science Research
Program (No. 2009-0086824) through the NRF of Korea.

\begin{deluxetable}{ccccccc}
\tablecaption{Target List.}
\scriptsize
\label{targets}
\tablewidth{17.0cm}
\startdata \\
\hline \hline
    CLUSTER      &     $FUV$ $t_{\rm exp}$ (sec)   &    $NUV$ $t_{\rm exp}$ (sec)   &
    ${\rm RA_{C}}$ (deg)   &   ${\rm Dec_{C}}$ (deg) & OBS Date & Cycle \\
\hline
 NGC~104   &      2235  &   4069   & 6.085   & -72.132   & 2006-07-06 & GI1  \\ 
 NGC~288   &      1606  &   1606   & 13.418  & -26.245   & 2004-12-06 & GI1/MIS \\ 
 NGC~362   &      2623  &   3027   & 15.809  & -70.848  & 2005-10-23 & GI1 \\ 
 NGC~1261  &      1225  &   1225   & 48.064  & -55.217  & 2004-12-09 & GI1 \\ 
 NGC~1851  &      2797  &   4487   & 78.526  & -40.047  & 2004-12-10 & GI1 \\ 
 NGC~1904  &      1326  &   3176   & 81.196  & -24.461  & 2004-12-14 & GI1 \\ 
 NGC~2298  &     10757  &  22171   & 102.066 & -35.945  & 2004-12-15 & GI1 \\
 NGC~2808  &       987  &    988   & 137.896 & -64.913  & 2007-03-11 & GI3 \\        
 NGC~2419  &      1262  &   3695   & 114.688 &  38.869 & 2008-12-16 & GI4	 \\ 
 NGC~4147  &      1678  &   1678   & 182.526 &  18.542  & 2006-03-29 & GI1 \\ 
 NGC~4590  &      1634  &   5081   & 190.020 & -26.605  & 2007-03-30 & GI1 \\ 
 NGC~5024  &      1656  &   1656   & 198.230 &  18.169  & 2007-05-02 & GI1 \\ 
 NGC~5053  &      1781  &   1782   & 199.112 &  17.698   & 2007-05-03 & GI1 \\ 
 NGC~5272  &      1679  &   1680   & 205.547 &  28.375  & 2007-05-01 & GI1 \\ 
 NGC~5466  &      1841  &   3532   & 211.364 &  28.535    & 2007-05-01 & GI1 \\ 
 NGC~5897  &      1590  &   2936   & 229.352 & -21.010  & 2007-06-06 & GI1 \\ 
 NGC~5904  &      1563  &   1566   & 229.592 &   2.069  & 2007-05-12 & GI3 \\    
 NGC~5986  &      4224  &   4225   & 236.514 &  -37.786 & 2007-06-06 & GI3 \\
 NGC~6101  &      2010  &   2010   & 247.039 & -72.502  & 2008-07-26 & GI1 \\ 
 NGC~6218  &       120  &  23891   & 251.811 & -1.948  & 2006-07-02 & GI1 \\ 
 NGC~6229  &      1603  &   5419   & 251.769 &  47.477  & 2007-04-13 & GI1 \\ 
 NGC~6235  &      1875  &  25131   & 253.373 & -22.585  & 2005-06-24 & GI1 \\ 
 NGC~6254  &      1911  &  25362   & 254.287 &  -4.099 & 2005-06-23 & GI1 \\ 
 NGC~6273  &      2264  &   ---    & 255.603 &  -26.563 & 2007-06-17 & GI3 \\
 NGC~6284  &      5767  &   4225   & 236.514 & -37.786  & 2007-06-06 & GI3 \\      
 NGC~6341  &      1911  &   1911   & 258.884 & 43.123  & 2008-05-25  & GI4 \\ 
 NGC~6342  &      3101  &   3101   & 260.730 &  -19.451 & 2008-05-27 & GI4  \\ 
 NGC~6402  &      5185  &   5184   & 264.500 &  -3.350  & 2007-06-17 & GI3  \\
 NGC~6356  &      3369  &   3369   & 260.949 & -17.642 & 2008-05-27 & GI4  \\ 
 NGC~6397  &      1584  &    409   & 265.574 &  -53.770& 2008-07-17 & GI4    \\
 NGC~6535  &      1671  &   1671   & 270.670 &  -0.330 & 2008-05-31 & GI4  \\ 
 NGC~6584  &      4799  &   4799   & 274.578 & -52.228 & 2008-07-17 & GI4 \\
 NGC~6809  &       840  &    840   & 294.994 & -31.063 & 2008-07-14 & GI4  \\  
 NGC~6864  &      1882  &   4817   & 301.520 & -21.921 & 2005-08-06 & GI1  \\ 
 NGC~6981  &      2470  &   5039   & 313.366 & -12.537 & 2005-08-05 & GI1  \\ 
\hline
\hline  
\enddata 
\end{deluxetable}

\setcounter{table}{0}
\begin{deluxetable}{ccccccc}
\tablecaption{Target List (continued).}
\scriptsize
\tablewidth{17.0cm}
\startdata \\
\hline \hline
    CLUSTER      &     $FUV$ $t_{\rm exp}$ (sec)   &    $NUV$ $t_{\rm
      exp}$ (sec)   &  ${\rm RA_{C}[deg]}$   &   ${\rm Dec_{C}[deg]}$ & OBS Date & Cycle\\
\hline
 NGC~7006  &      1457  &   4690   & 315.375 &  16.185 & 2006-08-12 & GI1    \\ 
 NGC~7089  &      3143  &   4418   & 323.372 & -0.823  & 2005-08-05 & GI1   \\ 
 NGC~7099  &      2305  &   2305   & 325.197 & -23.192 & 2008-08-04 & GI4   \\ 
 NGC~7492  &      1697  &   3302   & 347.224 & -15.639 & 2005-08-26   & GI1    \\      
 Arp~2     &      4027  &   4027   & 292.355    & -30.770    & 2008-07-11  & GI4   \\ 
 Pal~11    &      2120  &  15771   & 296.428    & -7.942     & 2005-06-17  & GI1  \\ 
 Pal~12    &      1510  &   3401   & 326.662    & -21.251 & 2006-08-01 & GI1     \\
 IC~4499   &      4279  &   4279   & 225.077    & -82.213 & 2008-07-29  & GI4   \\ 
 Terzan~8  &      3084  &   3084   & 295.438    & -34.000 & 2008-07-12  & GI4  \\     
\hline	  								    
\hline    
\enddata
\end{deluxetable}

\begin{deluxetable}{ccccccc}
\tablecaption{Standard stars used in the non-linearity tests.}
\scriptsize
\label{standards}
\tablewidth{15.cm}
\startdata \\
\hline \hline
   Star     &   $FUV$ & $FUV_{M07}$ & $FUV_{\rm predicted}$ & $NUV$ & $NUV_{M07}$ & $NUV_{\rm predicted}$ \\
\hline
 GD50    &12.74  &   12.70 &   11.98	&  12.82 &  12.84  &   12.57 \\    
 HZ4     &14.55  &   14.58 &   14.53	&  14.52 &  14.56  &   14.50  \\ 
 HZ2     &13.21  &   13.20 &   12.86	&  13.37 &  13.39  &   13.25  \\ 
G191B2B  &12.26  &   11.47 &   99.99	&  11.71 &  11.65  &   10.17  \\
GD108    &12.49  &   12.52 &   12.39	&  13.08 &  13.19  &   12.77  \\ 
HZ21     &13.11  &   12.99 &   12.55	&  13.27 &  13.30  &   13.13 	\\
GD153    &12.78  &   99.99 &   11.33	&  12.37 &  12.36  &   11.91  \\ 
HZ43     &12.73  &   12.31 &   10.75	&  11.98 &  11.98  &   11.36  \\  
LTT9491  &16.09  &   16.16 &   16.09	&  14.60 &  14.64  &   14.58  \\ 
G93      &12.94  &   99.99 &   12.14	&  12.66 &  12.67  &   12.39   \\ 
NGC7293  &12.12  &   10.03 &   10.93	&  12.38 &  99.99  &   11.70   \\ 
LDS749B  &15.63  &   15.66 &   15.57	&  14.75 &  14.78  &   14.71   \\ 
BD33     &12.87  &   12.35 &   10.51	&  12.75 &  12.66  &   10.47   \\       
\hline
\hline
\enddata

\end{deluxetable}

\begin{deluxetable}{lccccc}
\tablecaption{Post He-core burning candidates.}
\label{candidates}
\scriptsize
\tablewidth{13.0cm}
\startdata \\
\hline \hline
 STAR ID   &  $FUV$    & $NUV$  &  RA  &   Dec  & Class \\
\hline
 NGC~104-1  & 13.069  &  12.448  &   00:23:58.0  & -72:05:30  & P(E)AGB    \\ 
 NGC~288-5695  & 16.687  &  16.227 &   12:52:40.9  &  +26:33:53 & AGBM \\
 NGC~362-1625  & 14.038  &  14.885 &   1:03:11.5   &  -70:49:13   & AGBM \\
 NGC~362-372   & 13.649  &  14.496 &   1:02:16.2   & -70:51:42    & AGBM \\
 NGC~362-1626  & 13.772  &  14.632 &   1:03:38.8   & -70:49:12    & AGBM \\
 NGC~362-1444  & 13.769  &  14.791 &   1:01:53.1   & -70:54:11    & AGBM \\
 NGC~362-2413  & 13.262  &  14.412 &   1:03:12.2   & -70:59:39    & AGBM \\
 NGC~1261-1   & 14.670  &  15.483  &   3:11:48.6   &  -55:32:36  & AGBM \\
 NGC~1261-43   & 14.734  &  15.678  &   3:12:27.8   &  -55:34:59 & AGBM \\
 NGC~1261-8   & 14.307  &  15.376  &   3:11:56.2   &  -55:17:44   & AGBM \\
 NGC~1261-19   & 15.020  &  16.067 &   3:12:10.1   &  -55:12:22   & AGBM \\
 NGC~1851-44  & 10.895  &  12.355  &   05:14:08.6  & -40:03:03   & P(E)AGB \\
 NGC~2808-17  & 14.802  &  14.902  &   09:12:03.9  & -64:51:31 & P(E)AGB \\
 NGC~2808-69  & 15.824  &  15.874  &   09:12:02.1  & -64:52:36 & P(E)AGB \\
 NGC~2808-72  & 16.187  &  15.900  &   09:12:05.7  & -64:51:56 & P(E)AGB \\
 NGC~2808-84  & 16.295  &  16.026  &   09:12:07.2  & -64:51:37 & P(E)AGB \\
 NGC~2808-99  & 15.778  &  16.163  &   09:12:11.9  & -64:50:37 & P(E)AGB \\
 NGC~2808-103  & 16.178  &  16.193  &   09:12:05.4  & -64:52:37 & P(E)AGB \\
 NGC~2808-118  & 16.009  &  16.342  &   09:11:51.5  & -64:51:49 & P(E)AGB \\
 NGC~2808-119  & 16.235  &  16.344  &   09:12:06.1  & -64:50:35 & P(E)AGB \\
 NGC~2808-128  & 16.714  &  16.410  &   09:12:00.5  & -64:51:26 & P(E)AGB \\
 NGC~2808-132  & 16.404  &  16.424  &   09:12:06.5  & -64:52:00 & P(E)AGB \\
 NGC~2808-157  & 17.053  &  16.614  &   09:12:00.5  & -64:52:12 & P(E)AGB \\
 NGC~2808-173  & 16.614  &  16.753 &   9:12:01.7   &  -64:47:34   & AGBM \\
 NGC~2808-206  & 16.569  &  17.004 &   9:11:43.1   &  -64:52:29    & AGBM \\
 NGC~2808-254  & 17.152  &  17.229 &   9:12:22.1   &  -64:52:38     & AGBM \\
 NGC~2808-267  & 16.859  &  17.279 &   9:11:59.9   &  -64:53:26   & AGBM \\
 NGC~2808-314  & 17.185  &  17.477 &   9:12:25.4   &  -64:52:05   & AGBM \\
 NGC~2808-350  & 17.222  &  17.577 &   9:11:41.9   &  -64:42:06   & AGBM \\
 NGC~2808-387  & 18.110  &  17.702 &   9:12:05.3   &  -64:53:25   & AGBM \\
 NGC~2808-421  & 17.591  &  17.783 &   9:12:20.0   &  -64:50:55   & AGBM \\
 NGC~2808-594  & 18.439  &  18.216 &   9:12:05.5   &  -64:53:47   & AGBM \\
 NGC~2808-656  & 18.240  &  18.307 &   9:12:10.8   &  -64:50:02   & AGBM \\
 NGC~2808-670  & 18.024  &  18.328 &   9:11:24.9   &  -64:52:45   & AGBM \\
 NGC~2808-711  & 18.314  &  18.406 &   9:12:12.0   &  -64:53:15   & AGBM \\
 NGC~2808-756  & 18.288  &  18.482 &   9:11:50.9   &  -64:50:30   & AGBM \\
 NGC~2808-1146 &  18.255 &  18.944 &   9:12:17.2   &  -64:48:04   & AGBM \\
\hline
\hline
\enddata

\end{deluxetable}								         

\setcounter{table}{2}

\begin{deluxetable}{lccccc}
\tablecaption{Post He-core burning candidates (continued).}
\scriptsize
\tablewidth{13.0cm}
\startdata \\
\hline \hline
 STAR ID   &  $FUV$    & $NUV$  &  RA  &   Dec  & Class \\
\hline
 NGC~4590-9   & 15.925  &  15.627 &   12:38:33.0  &  -26:41:15   & AGBM \\
 NGC~5024-5    & 15.269  &  15.655  &   13:14:00.1 &  +18:31:31 & P(E)AGB \\
 NGC~5024-7    & 15.506  &  15.832  &   13:13:10.6 &  +18:07:36  & P(E)AGB \\
 NGC~5024-161  & 17.994  &  18.307 &   13:12:39.5  &  +18:04:54  & AGBM \\
 NGC~5272-2    & 13.613  &  13.778  &   13:42:16.9 &  +28:26:02   & P(E)AGB \\ 
 NGC~5272-32   & 16.137  &  16.488 &   13:42:01.2  & +28:23:25   & AGBM \\
 NGC~5272-38   & 16.146  &  16.527 &   13:42:05.9  & +28:19:05   & AGBM \\
 NGC~5466-1    & 13.075  &  13.997  &   14:03:17.2 &  +28:39:30 & PAGB \\
 NGC~5466-36   & 17.391  &  17.793 &   14:05:31.6  & +28:33:10      & AGBM \\
 NGC~5897-26   & 16.564  &  16.719 &   15:18:56.6  &  -21:08:40  & AGBM \\
 NGC~5897-34   & 17.578  &  17.080 &   15:17:23.7  & -20:37:56   & AGBM \\
 NGC~5897-50   & 17.308  &  17.648 &   15:18:04.7  &  -20:47:02  & AGBM \\
 NGC~5897-61   & 17.499  &  17.835 &   15:18:40.6  & -21:16:08  & AGBM \\
 NGC~5904-1    & 13.318  &  13.361  &   15:18:34.2 &  +02:05:02 & P(E)AGB \\
 NGC~5904-3   & 14.413  &  14.612 &   15:18:32.8  &  +01:54:54	   & AGBM \\
 NGC~5986-634  & 19.140  &  19.443 &   15:46:38.1  &   -37:41:25    & AGBM \\
 NGC~5986-701  &  19.165 &   19.580  &   15:46:46.5  & -37:49:38.     & AGBM \\
 NGC~6101-4874 &  16.782 &   17.159  &   16:26:47.1  & -72:15:24    & AGBM \\
 NGC~6218-48   &  16.643 &   17.423  &   16:47:12.6  &  -01:41:24    & AGBM \\
 NGC~6235-154  &  18.585 &   19.174  &   16:53:39.1  & -22:16:17  & AGBM \\
 NGC~6235-184  &  19.449 &   19.412  &   16:52:47.4  & -22:04:14    & AGBM \\
 NGC~6235-186  &  19.578 &   19.437  &   16:52:51.1  & -22:14:21  & AGBM \\
 NGC~6235-196  &  18.847 &   19.507  &   16:52:49.4  & -22:11:50    & AGBM \\
 NGC~6235-254  &  19.754 &   19.780  &   16:53:18.0  & -22:11:39   & AGBM \\
 NGC~6235-355  &  19.496 &   20.136  &   16:53:43.4  & -22:17:36   & AGBM \\
 NGC~6235-439  &  19.734 &   20.350  &   16:53:38.3  & -22:01:02   & AGBM \\
 NGC~6235-43   & 16.919  &  17.475  &   16:53:20.8 & -22:02:40  & P(E)AGB \\
 NGC~6254-66    & 14.244  &  15.544 &  16:56:48.0   &   -04:04:33    & AGBM \\
 NGC~6254-112  &  13.813 &   15.052  &   16:57:02.8  &  -04:08:19    & AGBM \\
 NGC~6254-117  &  13.990 &   15.071  &   16:57:05.2  & -04:07:56   & AGBM \\
 NGC~6254-152   & 13.208  &  14.829 &  16:56:43.7   &  -04:05:41    & AGBM \\
 NGC~6254-189   & 14.390  &  15.851 &  16:57:06.3   &  -04:03:19     & AGBM \\
 NGC~6254-241   & 13.433  &  14.564 &  16:57:14.7   &  -04:05:03     & AGBM \\
 NGC~6254-242  & 11.096  &  12.905  &   16:57:09.4 & -04:04:24 & P(E)AGB \\
 NGC~6254-364   & 14.167  &  15.680 &  16:57:01.1   &  -04:04:30     & AGBM \\
 NGC~6284-2    & 13.175  &  13.772  &   17:04:10.4 & -24:27:57 & P(E)AGB \\
 NGC~6284-85   & 16.156  &  16.484  &   17:04:29.7 & -24:29:20 & P(E)AGB \\
\hline
\hline
\enddata

\end{deluxetable}								         

\setcounter{table}{2}

\begin{deluxetable}{lccccc}
\tablecaption{Post He-core burning candidates (continued).}
\scriptsize
\tablewidth{13.0cm}
\startdata \\
\hline \hline
 STAR ID   &  $FUV$    & $NUV$  &  RA  &   Dec  & Class \\
\hline
 NGC~6284-116  & 17.032  &  16.921  &   17:04:45.0 & -24:32:60 & P(E)AGB \\
 NGC~6284-154  & 17.282  &  17.207  &   17:05:10.8 & -24:32:24 & P(E)AGB \\
 NGC~6284-212  & 17.385  &  17.606  &   17:03:11.4 & -24:51:31 & P(E)AGB \\
 NGC~6342-110  & 18.991  &  18.803  &   17:21:38.3 & -19:34:04 & P(E)AGB \\
 NGC~6356-1    & 13.089  &  14.059  &   17:23:25.2 & -17:58:15 & P(E)AGB \\
 NGC~6356-311   & 18.480  &  18.867 &  17:23:45.9   &  -17:41:59 & AGBM \\
 NGC~6356-424   & 18.877  &  19.243 &  17:24:04.0   &  -17:49:37 & AGBM \\
 NGC~6356-480   & 19.357  &  19.390 &  17:23:45.6   & -17:50:17     & AGBM \\
 NGC~6356-849   & 19.745  &  20.061 &  17:23:39.9  &  -17:43:57 & AGBM \\
 NGC~6397-149   & 14.773   & 15.029  &  17:41:30.620 & -53:28:18.90 & AGBM \\
 NGC~6397-438   & 14.827  &  14.330  &  17:39:44.524 & -53:43:29.37 & AGBM \\
 NGC~6397-522   & 13.640  &  13.680  &  17:40:38.428 & -53:38:32.20 & AGBM \\
 NGC~6402-31   & 17.581  &  17.508  &   17:37:33.2 & -03:14:52 & PAGB \\
 NGC~6402-58   & 18.079  &  18.528  & 	17:37:37.3 & -03:15:45 & P(E)AGB \\
 NGC~6402-92   & 19.206  &  19.170  & 	17:37:38.1 & -03:14:09 & P(E)AGB \\
 NGC~6402-99   & 19.142  &  19.261  & 	17:37:33.6 & -03:15:27 & P(E)AGB \\
 NGC~6402-102  & 18.669  &  19.321  & 	17:38:20.1 & -03:10:06 & P(E)AGB \\
 NGC~6402-142  & 19.169  &  19.718  & 	17:37:26.1 & -03:14:55 & P(E)AGB \\
 NGC~6402-143  & 19.045  &  19.718  & 	17:37:28.6 & -03:15:17 & P(E)AGB \\
 NGC~6402-156  & 19.663  &  19.823  & 	17:37:30.9 & -03:17:40 & P(E)AGB \\
 NGC~6402-160  & 19.039  &  19.838  & 	17:36:38.0 & -03:23:12 & P(E)AGB \\
 NGC~6402-171  & 19.627  &  19.900  & 	17:37:37.2 & -03:14:60 & P(E)AGB \\
 NGC~6402-193  & 20.046  &  20.050  & 	17:37:36.4 & -03:15:34 & P(E)AGB \\
 NGC~6402-202  & 20.016  &  20.074  & 	17:37:31.8 & -03:15:01 & P(E)AGB \\
 NGC~6402-224  & 20.167  &  20.144  & 	17:37:40.5 & -03:14:58 & P(E)AGB \\
 NGC~6864-8    & 15.853  &  15.412  & 	20:06:05.5 & -21:54:59 & PAGB \\
 NGC~6864-52   & 16.892  &  17.383  & 	20:05:51.3 & -21:42:19 & P(E)AGB \\
 NGC~6864-77    & 18.125  &  17.832 &  20:05:19.2   & -22:04:21 & AGBM \\
 NGC~6864-102   & 18.669  &  18.126 &  20:07:01.8   & -21:46:18 	 & AGBM \\
 NGC~6864-212   & 19.620  &  19.086 &  20:04:55.2   & -21:51:57 	  & AGBM \\
 NGC~6864-224   & 19.632  &  19.138 &  20:06:40.8 &  -22:00:23  & AGBM \\
 NGC~6864-225   & 19.631  &  19.140 &  20:05:21.5 &   -21:53:06   & AGBM \\
 NGC~6864-452   & 19.286  &  20.017 &  20:06:10.5 &  -21:38:58	 & AGBM \\
 NGC~7006-16    & 16.834  &  17.385 &  21:01:35.1 &  +16:06:10	 & AGBM \\
 NGC~7089-407  & 12.492  &  13.221  & 	21:33:31.4 & -00:49:09  & P(E)AGB \\
 NGC~7089-387  & 12.839  &  17.126  & 	21:33:35.6 &  -00:51:22 & P(E)AGB \\
 NGC~7089-234   & 14.175  &  15.013 &  21:33:19.7 & -00:47:5 & AGBM \\
 NGC~7089-194   & 14.788  &  15.834 &  21:33:17.9 &  -00:49:58 & AGBM \\
 NGC~7089-765   & 14.326  &  14.938 &  21:32:29.4 & -00:48:31 & AGBM \\
\hline
\hline
\enddata

\end{deluxetable}

\setcounter{table}{2}

\begin{deluxetable}{lccccc}
\tablecaption{Post He-core burning candidates (continued).}
\scriptsize
\tablewidth{13.0cm}
\startdata \\
\hline \hline
 STAR ID   &  $FUV$    & $NUV$  &  RA  &   Dec  & Class \\
\hline
 NGC~7089-89    & 14.571  &  15.579 &  21:33:30.9 &  -00:47:20 & AGBM \\
 NGC~7099-2    & 13.449  &  13.362  & 	21:39:56.7 &  -23:11:56 & P(E)AGB \\
 NGC~7099-10    & 14.763  &  15.058 &  21:40:18.1 & -23:13:23 & AGBM \\
 NGC7~099-68    & 16.545  &  16.651 &  21:41:38.7 & -22:54:11 & AGBM \\
 Arp~2-25      & 18.227  &  17.757  & 	19:27:45.3 & -30:24:51 & P(E)AGB \\
 Arp~2-61       & 19.555  &  19.860 &  19:29:11.2 &  -30:17:09  & AGBM \\
 Arp~2-70       & 19.831  &  20.254 &  19:28:57.0 &  -30:27:59  & AGBM \\
 Pal~12-25      & 16.423  &  16.498 &  21:47:49.2  &  -21:17:33 & AGBM \\
 Pal~12-59      & 18.094  &  17.556 &  21:46:04.3  &  -21:21:46 & AGBM \\
 IC~4499-485    & 19.811  &  20.409 &  14:56:54.2  & -82:12:22	 & AGBM \\
 Ter~8-38      & 17.800  &  17.270  & 	19:41:40.8 & -34:03:59& P(E)AGB \\
\hline						    
\hline  					    
\enddata 					    
\end{deluxetable}

\newpage

\newpage

\newpage

\newpage

\newpage

\clearpage

\setcounter{figure}{10}
\begin{figure}[!hp]
\begin{center}
\includegraphics[scale=0.9]{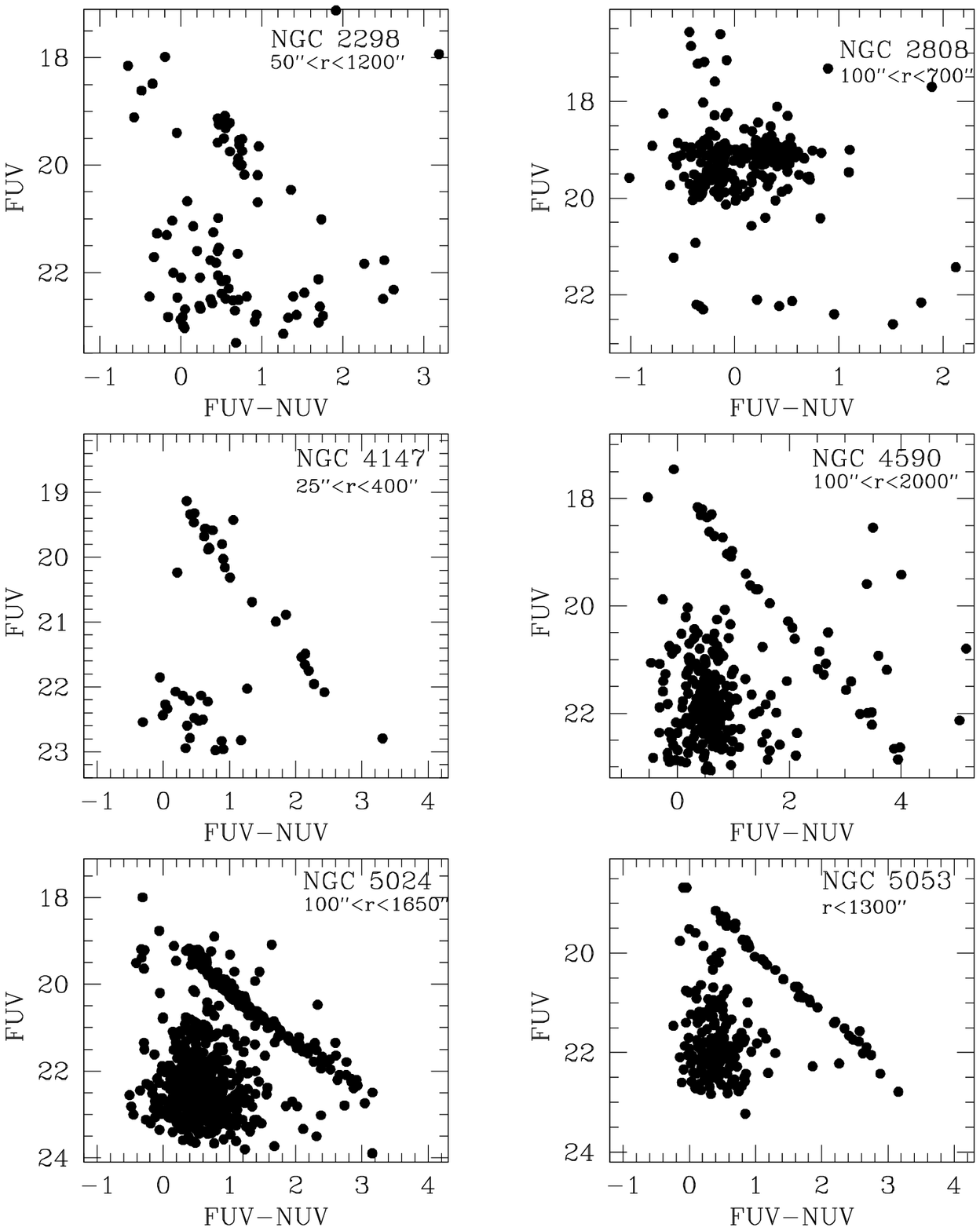}
\caption {b. Color-magnitude diagrams of Galactic GCs, continued.
The paucity of sources in the CMD of NGC~2808 is due to the shallowness
of the exposures for this cluster. }
\label{to2}
\end{center}
\end{figure}

\setcounter{figure}{10}
\begin{figure}[!hp]
\begin{center}
\includegraphics[scale=0.9]{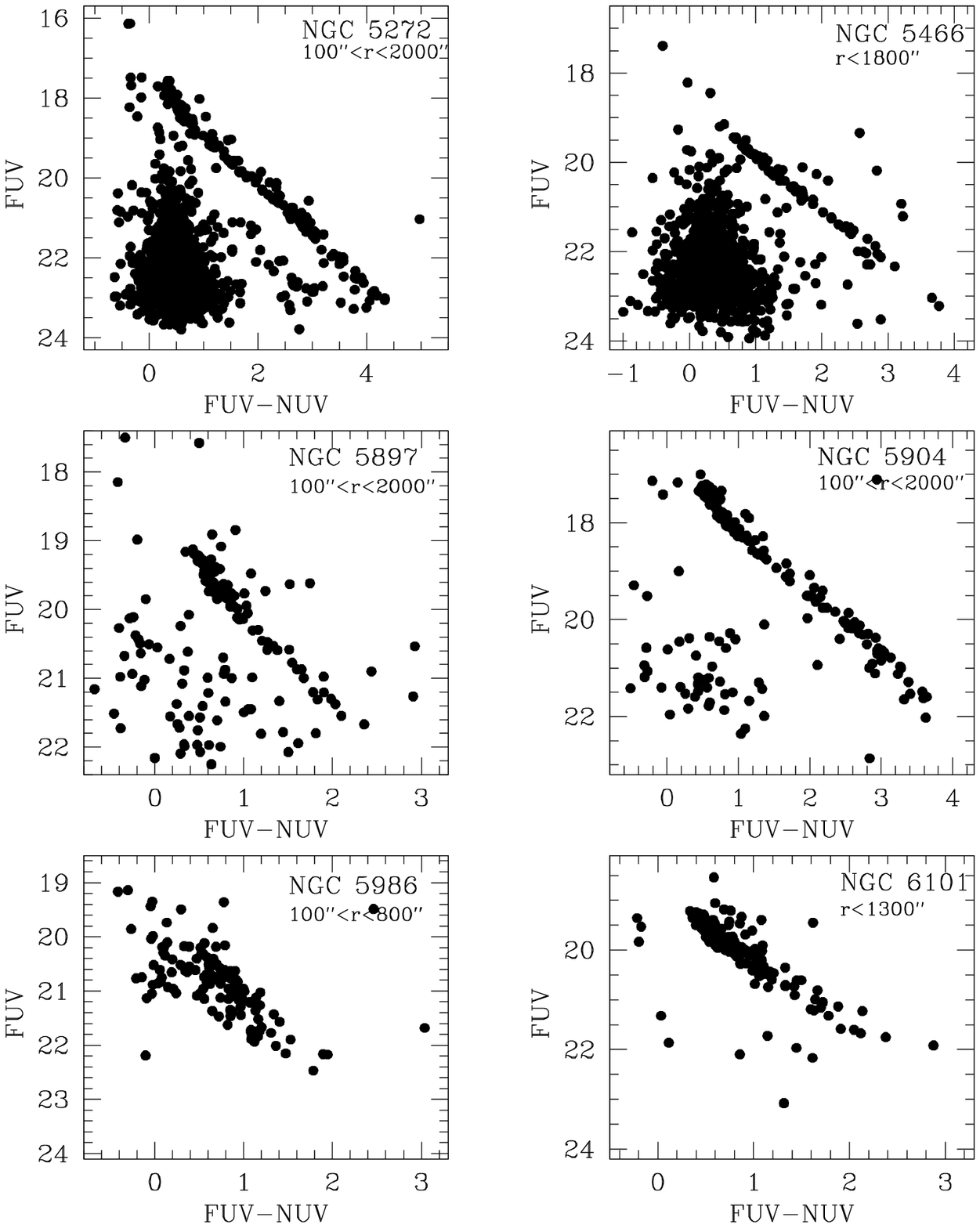}
\caption{c. Color-magnitude diagrams of Galactic GCs, continued.}
\label{to3}
\end{center}
\end{figure}

\setcounter{figure}{10}
\begin{figure}[!hp]
\begin{center}
\includegraphics[scale=0.9]{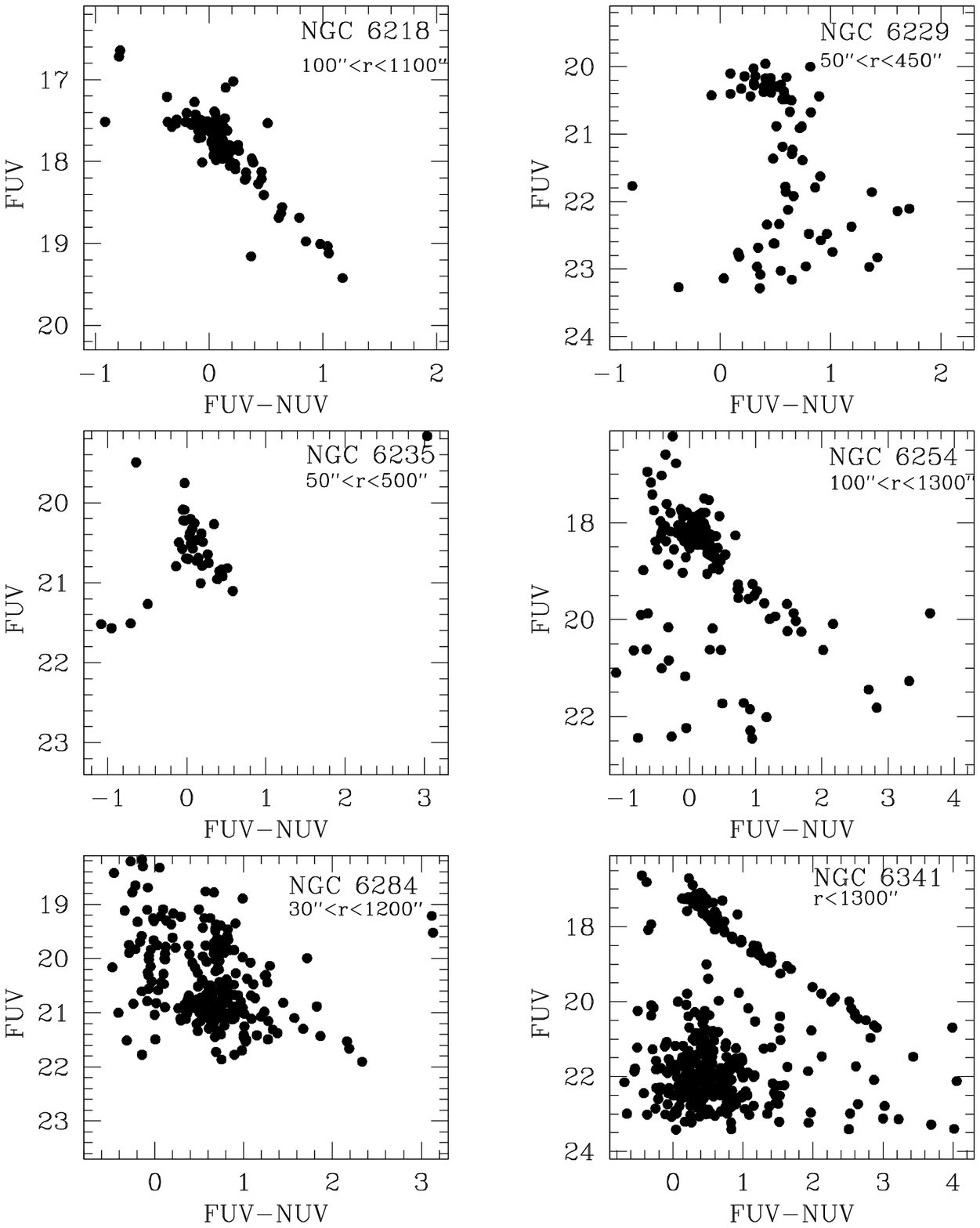}
\caption{d. Color-magnitude diagrams of Galactic GCs, continued.}
\label{to4}
\end{center}
\end{figure}

\setcounter{figure}{10}
\begin{figure}[!hp]
\begin{center}
\includegraphics[scale=0.9]{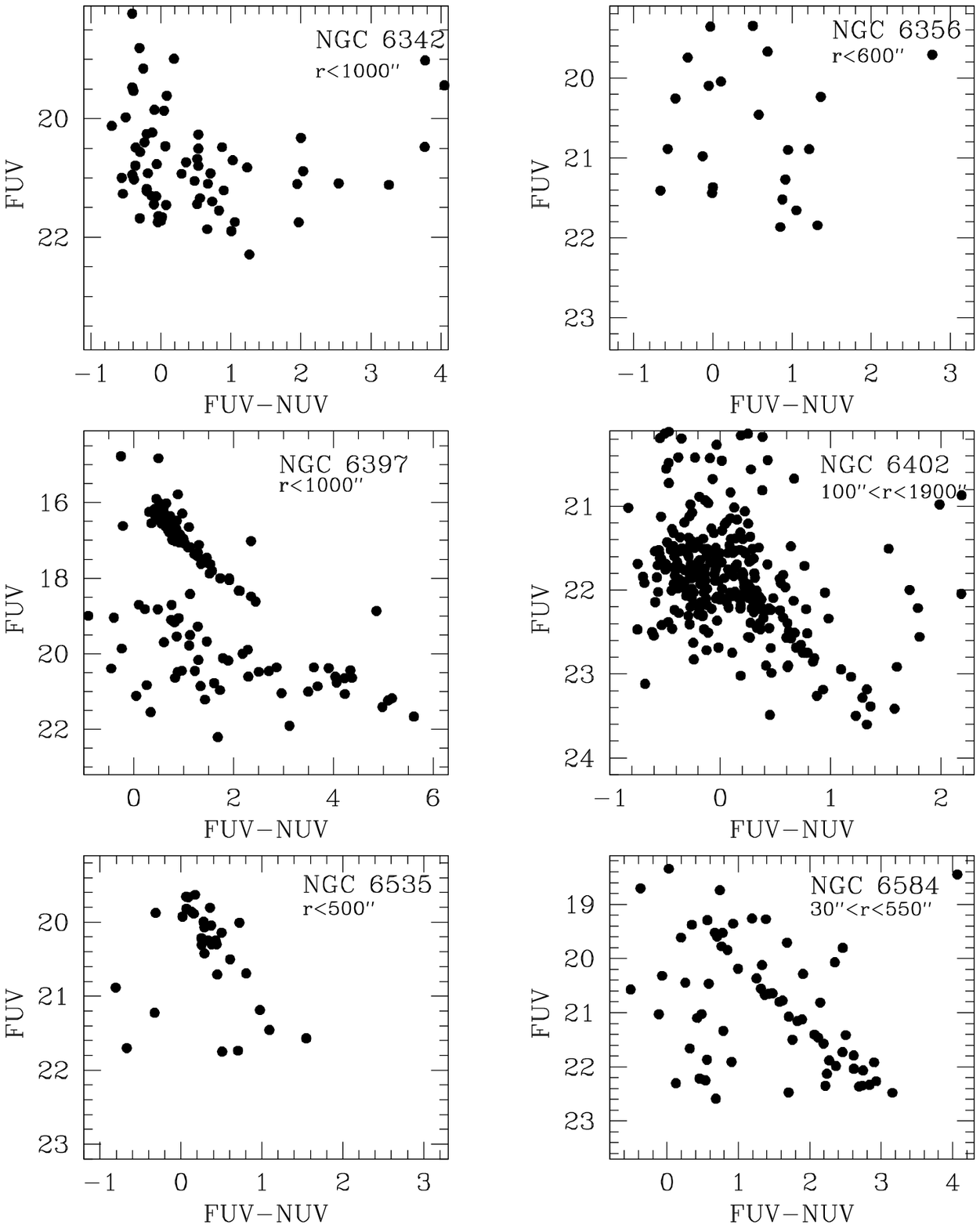}
\caption{e. Color-magnitude diagrams of Galactic GCs, continued.}
\label{to5}
\end{center}
\end{figure}

\setcounter{figure}{10}
\begin{figure}[!hp]
\begin{center}
\includegraphics[scale=0.9]{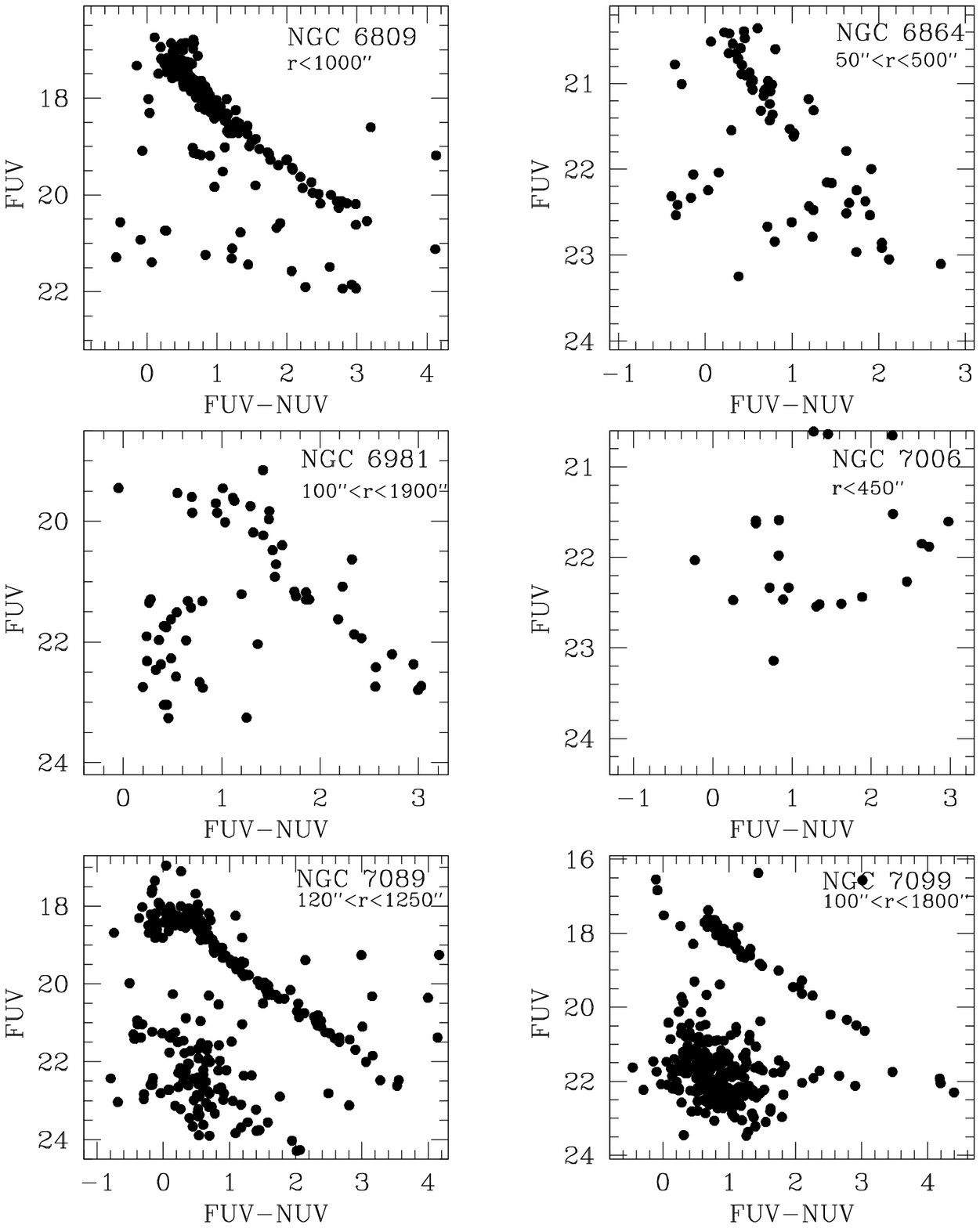}
\caption{f. Color-magnitude diagrams of Galactic GCs, continued.}
\label{to6}
\end{center}
\end{figure}

\setcounter{figure}{10}
\begin{figure}[!hp]
\begin{center}
\includegraphics[scale=0.9]{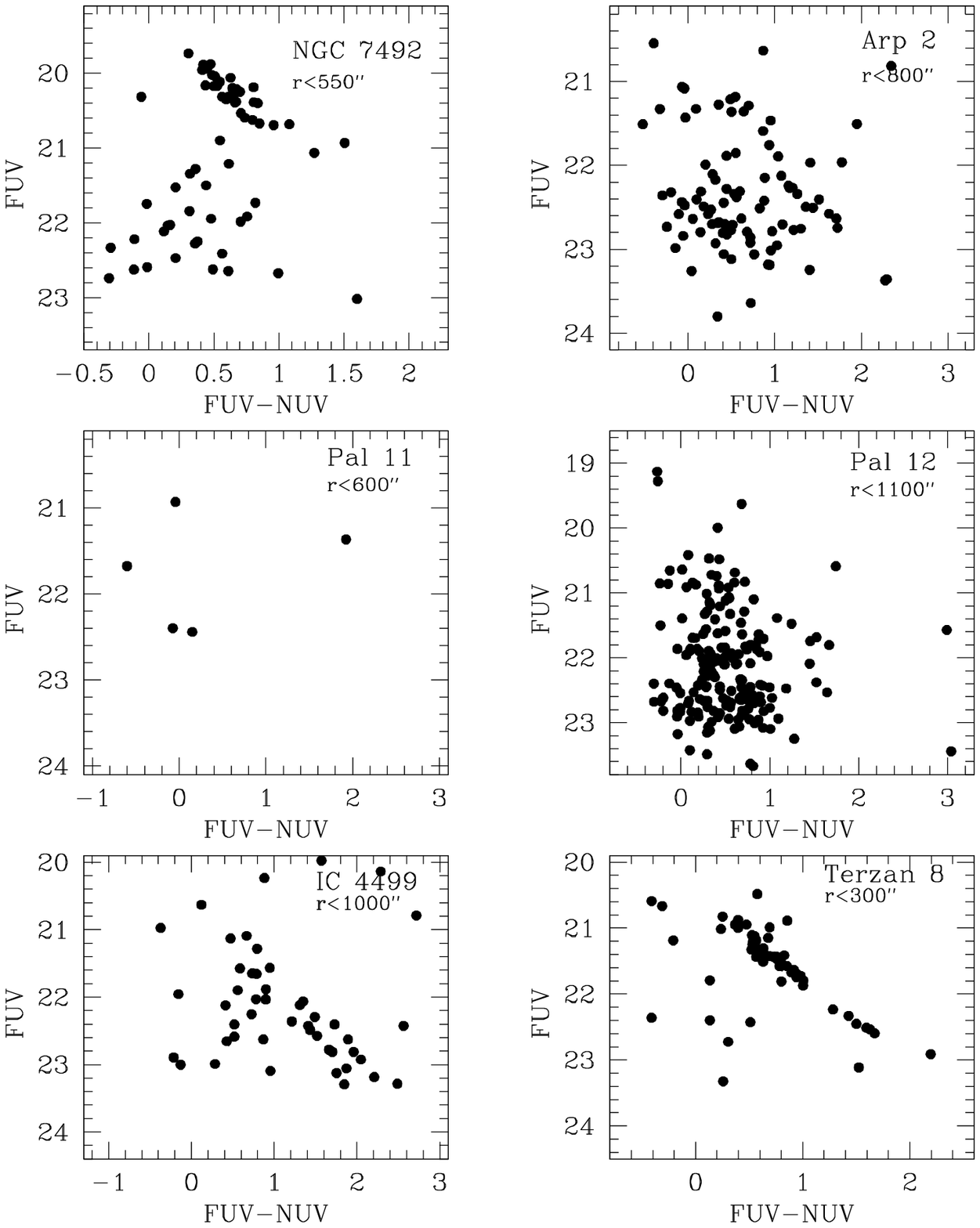}
\caption{g. Color-magnitude diagrams of Galactic GCs, continued.}
\label{to7}
\end{center}
\end{figure}

\begin{figure}[!hp]
\begin{center}
\includegraphics[scale=0.7]{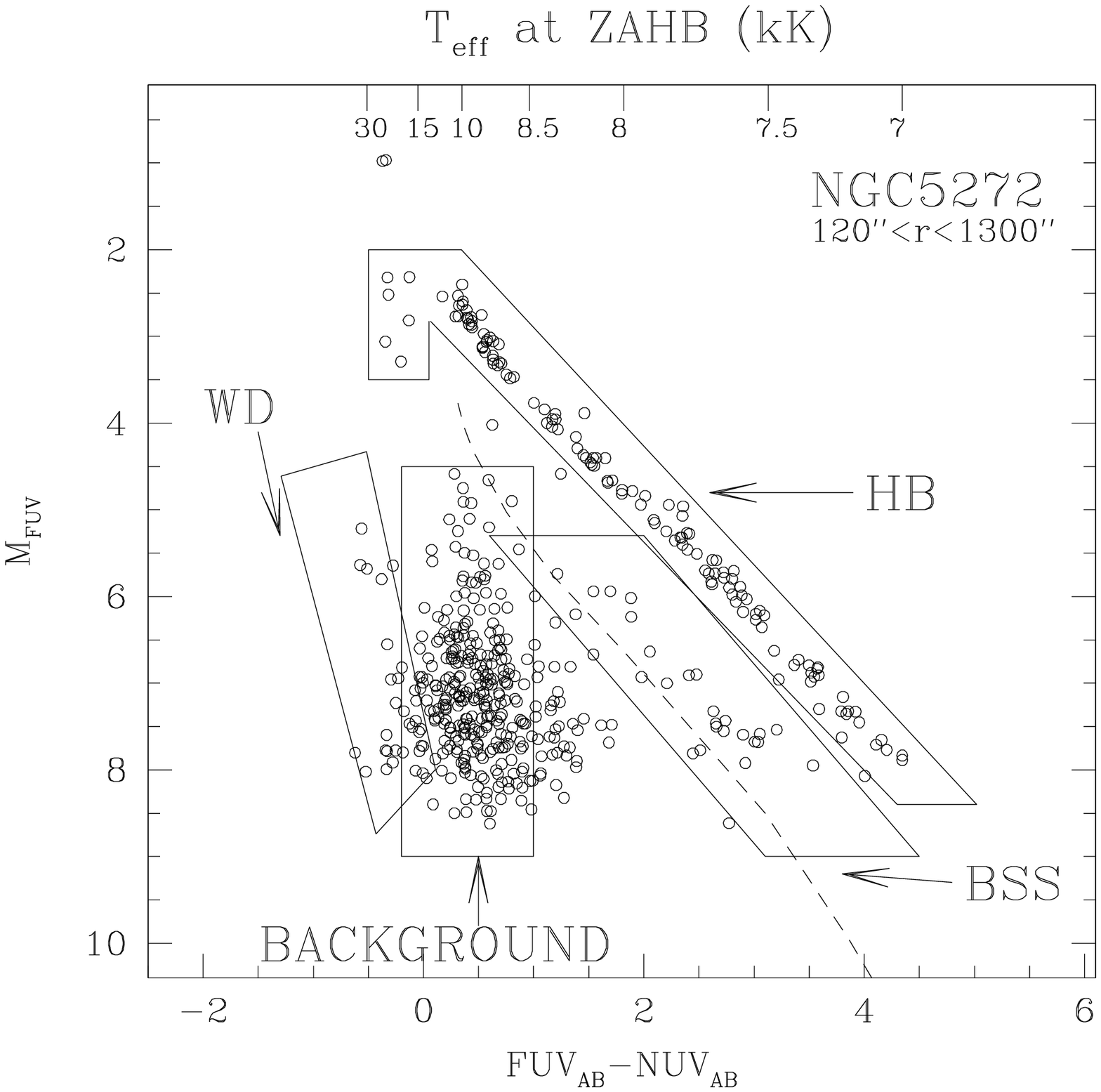}
\caption{Reddening- and distance modulus-corrected CMD of NGC~5272
(M~3), indicating the main populations that dominate the UV light
of old stellar populations.  The $T_{\rm eff}$ scale on the top
axis was obtained using Kurucz model fluxes
(http://kurucz.harvard.edu/grids.html) for the cluster metallicity,
adopting surface gravities from \cite{do93} and is appropriate only
for HB stars.  Note the presence of gaps in the cluster horizontal
branch, and an artificial clump of HB stars around $(FUV - NUV)
\sim 0.5$, which is due to the strongly non-linear color-$T_{\rm
eff}$ relation.  The brightest and hottest cluster blue stragglers
are clearly detected, 1--1.5~mag below the HB.  Two candidate PAGB
stars are visible at $M_{\rm FUV} \sim 1$ and $(FUV - NUV) \sim
-0.5$.  A few white dwarf candidates are also detected, but, at the
GALEX resolution, it is very difficult to distinguish them from
background sources.  The latter are predominantly extragalactic.
}
\label{5272}
\end{center}
\end{figure}

\begin{figure}[!hp]
\begin{center}
\includegraphics[scale=0.7]{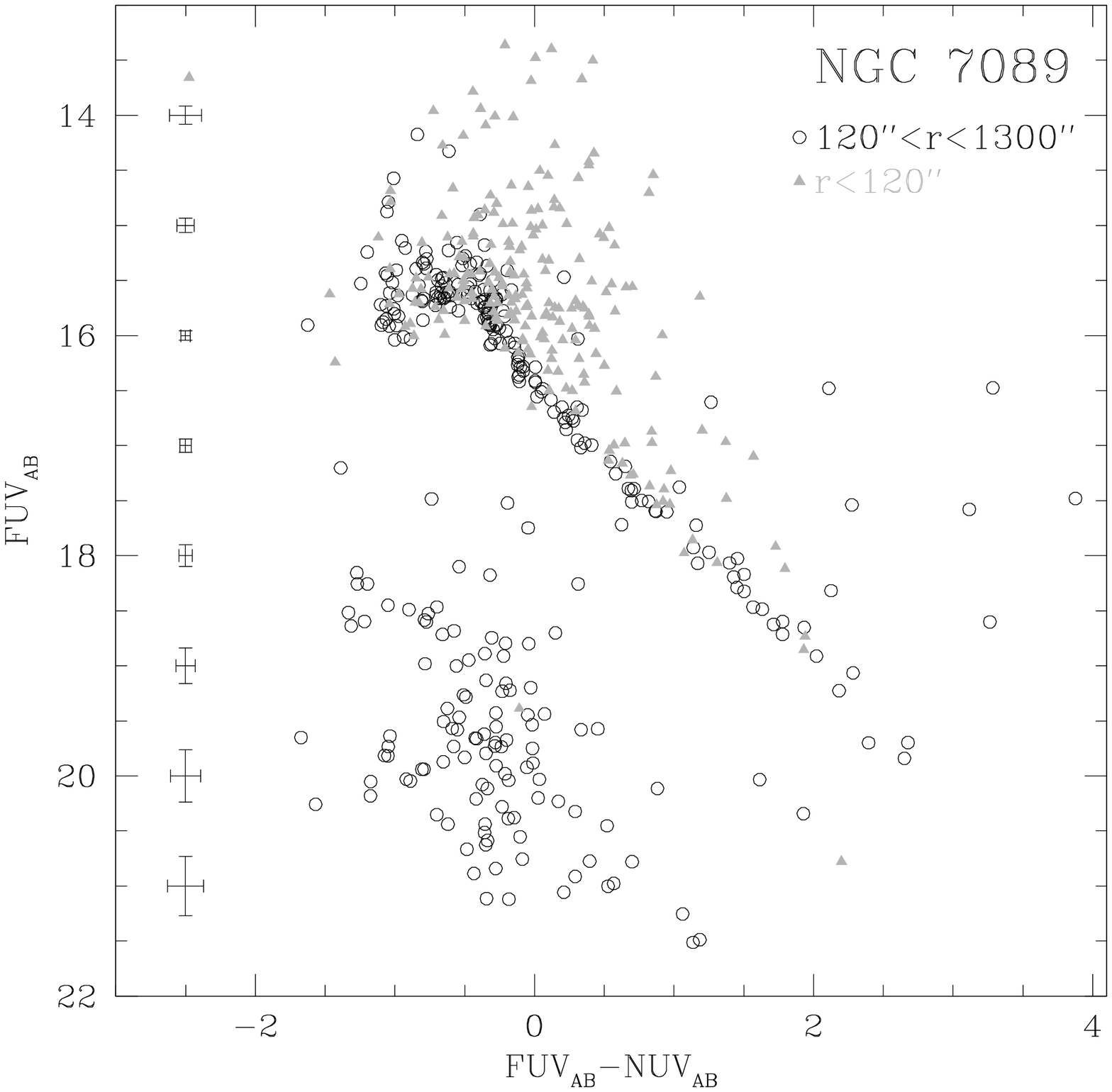}
\caption{Observed CMD of one of our densest clusters, NGC~7089,
illustrating the effect of crowding in our photometry.  Gray triangles
represent sources within 2$\arcmin$ of the cluster center, and all
others are located between that inner radial distance and the cluster
tidal radius (1300$\arcsec$).  The main effect of crowding is to
displace stars towards brighter $FUV$ magnitudes and redder colors.
The color effect is due to the fact that crowding is more sever in
the $NUV$ than in the $FUV$.
}
\label{7089}
\end{center}
\end{figure}

\begin{figure}[!hp]
\begin{center}
\includegraphics[scale=0.9]{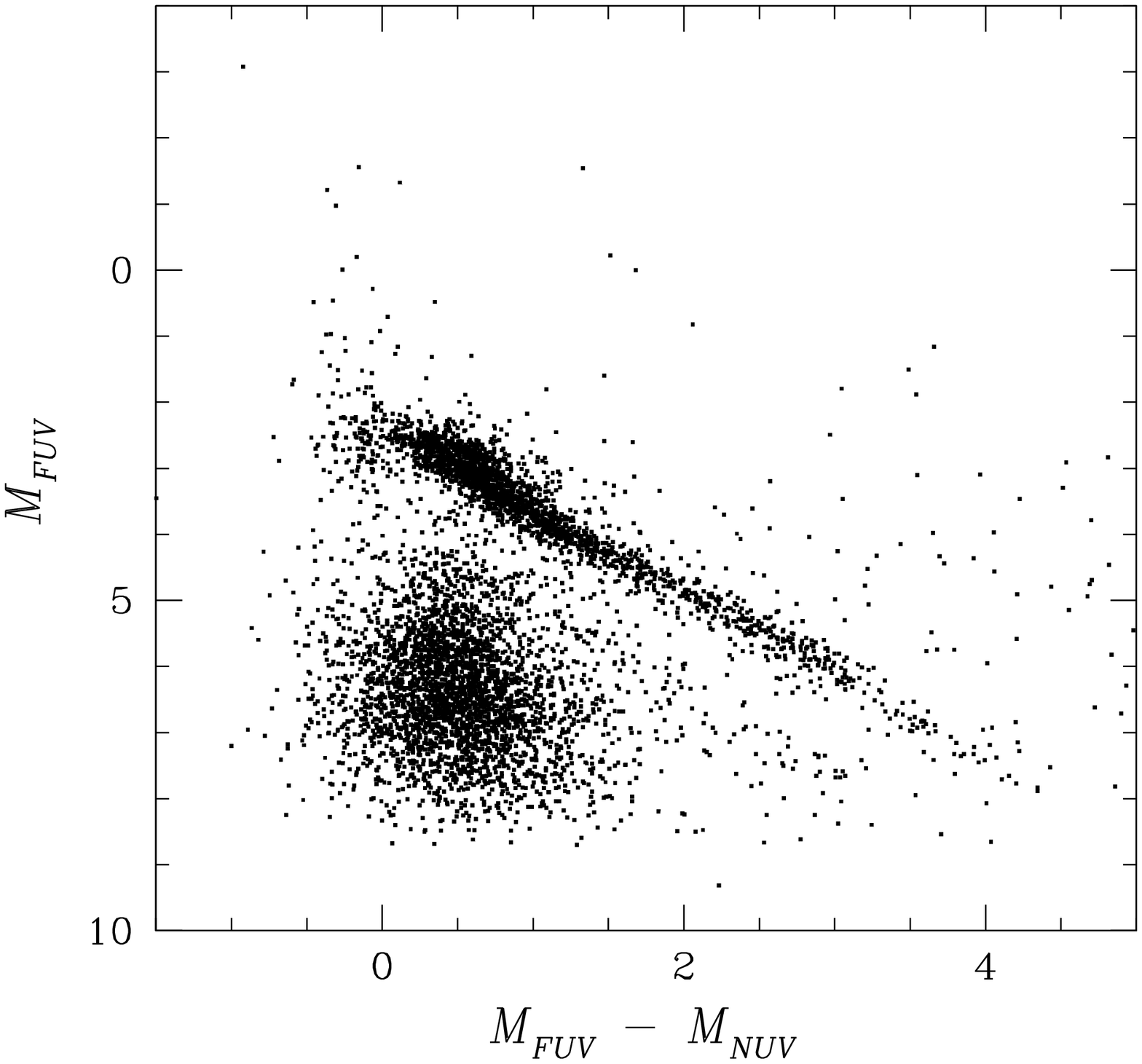}
\caption{Stacked color-magnitude diagram of 23 Galactic GCs (see Sect.~4). The UV-bright
population composed by candidate post He-core burning at $(M_{FUV}- M_{NUV}) \sim\,0$
is clearly seen in this stack.  Redder
stars brighter than the HB are likely to be predominantly background
sources.  The blue stragglers can also be very clearly spotted in this
diagram, below the redder half of the HB.}
\label{pagb_data}
\end{center}
\end{figure}

\begin{figure}[!hp]
\begin{center}
\includegraphics[scale=0.9]{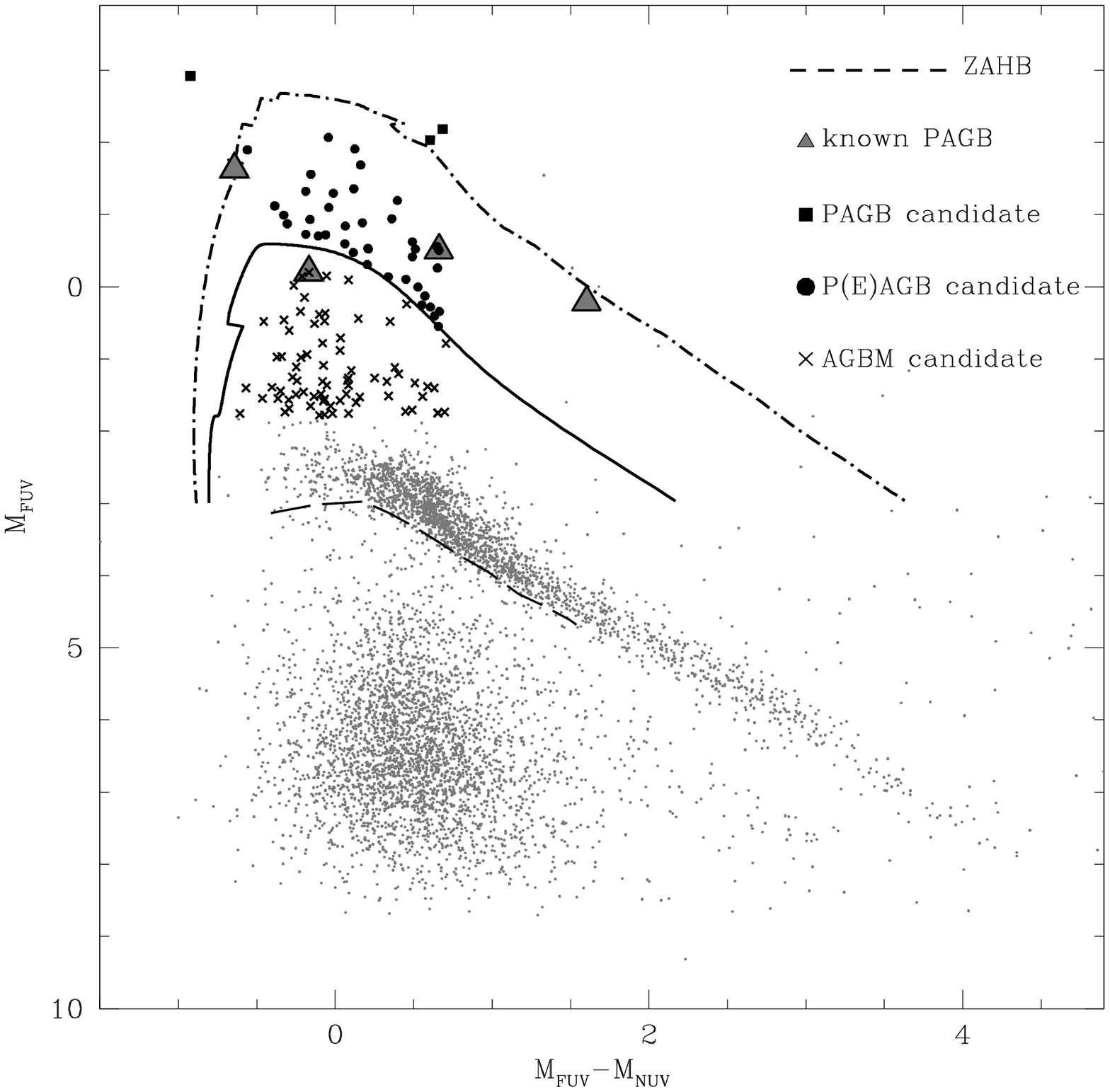}
\caption{Stacked color magnitude diagram from Figure~\ref{pagb_data}
(gray dots), with theoretical models from \cite{br08} overlayed on
the data.  The thick solid line represents the post-HB evolutionary
path for a star of $M=0.77M_{\odot}$, while the dashed-dotted line
is for a $M=0.515M_{\odot}$ star. The dashed model is the ZAHB.
The solid squares represent candidate PAGB stars. Large circles are
candidate P(E)AGBs and crosses are candidate AGBM stars. Large gray
triangles indicate the positions of a few well-known PAGB stars.
Note that, while for all the other stellar types photometry is only
plotted for stars within the cluster-centric limits shown in
Figures~\ref{to1} to \ref{to7}, photometry is shown for all PAGB
stars.  See discussion in Section~\ref{uvbright}.  }
\label{pagb_mod}
\end{center}
\end{figure}

{}


\begin{thebibliography}{}

\bibitem[Borkova \& Marsakov (2000)]{bm00} Borkova, T.V. \& Marsakov, V.A.
2000, Astronomy Reports, 44, 665
\bibitem[Brown \etal (2001)]{br01} Brown, T.M., Sweigart, A.V., Lanz, T.,
	Landsman, W.B. \& Hubeny, I. 2001, \apj, 562, 638
\bibitem[Brown \etal (2008)]{br08} Brown, T.M., Smith, E., Ferguson, H.C.,
	Sweigart, A.V., Kimble, R.A> \& Bowers, C.W. 2008, apj, 682
\bibitem[Busso \etal (2007)]{bu07} Busso, G., Cassisi, S., Piotto, G. \etal
2007, \aap, 474, 105
\bibitem[Cardelli \etal (1989)]{ca89} Cardelli, J.A., Clayton, G.C. \&
Mathis, J.S. 1989, \apj, 345, 245
\bibitem[Catelan (2008)]{ca08} Catelan, M. 2008, Mem. Soc. Astron.
Italiana, 79, 388
\bibitem[Catelan (2009)]{ca09} Catelan, M. 2009, \apss, 320, 261
\bibitem[Code (1969)]{co69} Code, A.D. 1969, \pasp, 81, 475
\bibitem[Dalessandro et al.(2008)]{da08} Dalessandro, E., 
Lanzoni, B., Ferraro, F.~R., et al.\ 2008, \apj, 681, 311 
\bibitem[Dalessandro \etal (2009)]{da09} Dalessandro, E., Beccari, G.,
	Lanzoni, B., Ferraro, F.R., Schiavon, R. \& Rood, R.T. 2009, \apjs,
	182, 509
\bibitem[Dalessandro \etal (2011)]{da11} Dalessandro, E. Salaris, M.,
	Ferraro, F.R. \etal, \mnras, 410, 694
\bibitem[Dorman \etal (1993)]{do93} Dorman, B., Rood, R.T. \& O'Connell,
	R.W. 1993, \apj, 419, 596
\bibitem[Dorman \etal (1995)]{do95} Dorman, B., O'Connell, R.W. \& Rood,
	R.T. 1995, \apj, 442, 105
\bibitem[Ferraro et al.(1997)]{fe97} Ferraro, F.~R., 
	Paltrinieri, B., Fusi Pecci, F., et al.\ 1997, \aap, 324, 915	
\bibitem[Ferraro \etal (1998)]{fe98} Ferraro, F.R., Paltrinieri, B., Fusi
Pecci, F., Rood, R.T. \& Dorman, B. 1998, \apj, 500, 311
\bibitem[Ferraro \etal (1999)]{fe99} Ferraro, F.R. , Paltrinieri, B., Rood,
	R.T. \& Dorman, B. 1999, \apj, 522, 983
\bibitem[Ferraro \etal (2001)]{fe01} Ferraro, F.R., D'Amico, N., Possenti,
	A., Mignani, R.P. \& Paltrinieri, B. 2001, \apj, 561, 337
\bibitem[Ferraro \etal (2003)]{fe03} Ferraro, F.R., Sills, A., Rood, R.T.,
	Paltrinieri, B. \& Buonanno, R. 2003, \apj, 588, 464
\bibitem[Girardi \etal (2000)]{gi00} Girardi, L., Bressan, A., Bertelli,
	G. \& Chiosi, C. 2000, \aaps 141, 371
\bibitem[Greggio \& Renzini (1990)]{gr90} Greggio, L. \& Renzini, A. 1990,
	\apj, 364, 35
\bibitem[Greggio \& Renzini (1999)]{gr99} Greggio, L. \& Renzini, A. 1999,
	MmSAI, 70, 691
\bibitem[Grundahl \etal (1998)]{gru98} Grundahl, F., Vandenberg, D.A. \&
Andersen, M.I. 1998, \apj, 500, L179
\bibitem[Grundahl \etal (1999)]{gru99} Grundahl, F., Catelan, M., Landsman,
	W.B., Stetson, P.B. \& Andersen, M.I. 1999, \apj, 524, 242.
\bibitem[Harris (1996)]{ha96} Harris, W.E. 1996, \aj, 112, 1487
\bibitem[Hill \etal (1992)]{hi92} Hill, R.S. \etal\ 1992, \apjl, 395. L17
\bibitem[Landsman \etal (1996)]{la96} Landsman, W.~B.,
      Sweigart, A.~V., Bohlin, R.~C., et al.\ 1996, \apjl, 472, L93
\bibitem[Lanzoni \etal (2007)]{la07} Lanzoni, B., Sanna, N., Ferraro, F.R.,
	Valenti, E., Beccari, G., Schiavon, R.P., Rood, R.T., Mapelli, M.
	\& Sigurdsson, S. 2007, \apj, 663, 1040
\bibitem[Law \& Majewski (2010)]{lm10} Law, D.R. \& Majewski, S.R. 2010,
	\apj, 718, 1128
\bibitem[Lee \etal (1994)]{le94} Lee, Y.-W., Demarque, P. \& Zinn, R. 1994,
	\apj, 423, 248
\bibitem[Lee \etal (2005)]{le05} Lee, Y.-W., Joo, S.-J., Han, S.-I. \etal
2005, \apj, 621, L57
\bibitem[Moehler (2001)]{mo01} Moehler, S. 2001, \pasp, 113, 162
\bibitem[Moehler \etal (2004)]{mo04} Moehler, S., Sweigart, A.V., Landsman,
	W.B., Hammer, N.J. \& Dreizler, S. 2004, \aap, 415, 313
\bibitem[Morrisey \etal (2005)]{mo05} Morrissey, P. \etal\ 2005, \apj, 619,
	L7
\bibitem[Morrisey \etal (2007)]{mo07} Morrissey, P. \etal\ 2007, \apjs, 173,
	682
\bibitem[O'Connell (1999)]{oc99} O'Connell, R.W. 1999, \araa, 37, 603
\bibitem[O'Connell \etal (1997)]{oc97} O'Connell, R.W. 1997, Dorman, B.,
	Shah, R.Y., Rood, R.T., Landsman, W.B., Witt, A.N., Bohlin, R.C.,
	Neff, S.G., Roberts, M.S., Smith, A.M. \& Stecher, T.P.  1997, \aj,
	114, 1982
\bibitem[Parise \etal (1994)]{pa94} Parise, R.A. \etal\ 1994, \apj, 423, 305
\bibitem[Rey \etal (2007)]{re07} Rey, S.-C. \etal\ 2007, \apjs, 173, 643
\bibitem[Rey \etal (2009)]{re09} Rey, S.-C. \etal\ 2009, \apj, 700, L11
\bibitem[Rood \etal (2008)]{ro08} Rood, R.T., Beccari, G., Lanzoni, B.,
	Ferraro, F.R., Dalessandro, E. \& Schiavon, R.P. 2008, Mem. Soc.
	Astron. Italiana, 79, 283
\bibitem[Rood \etal (2011)]{ro11} Rood, R.T. \etal\ 2011, in preparation
\bibitem[Schiavon (2007)]{s07} Schiavon, R.P. 2007, \apjs, 171, 146
\bibitem[Sohn \etal (2006)]{so06} Sohn, S.T., O'Connell, R.W., Kundu, A.,
	Landsman, W.B., Burstein, D., Bohlin, R., Frogel, J.A. \& Rose,
	J.A. 2006, \aj, 131, 866
\bibitem[Sohn \etal (2011)]{ro11} Sohn, S. T.. \etal\ 2011, in preparation	
\bibitem[Stetson (1987)]{st87} Stetson, P.B. 1987, \pasp, 99, 191
\bibitem[Stetson et al.(1989)]{1989AJ.....97.1360S} Stetson, P.~B., Hesser, 
J.~E., Smith, G.~H., Vandenberg, D.~A., \& Bolte, M.\ 1989, \aj, 97, 1360 
\bibitem[Trager \etal (2005)]{tr05} Trager, S.C., Worthey, G., Faber, S.M.
\& Dressler, A. 2005, \mnras, 362, 2
\bibitem[van Winckel (2003)]{vw03} van Winckel, H. 2003, \araa, 41, 391
\bibitem[Whitney \etal (1994)]{wh94} Whitney, J.H. \etal\ 1994, \aj, 108,
	1350


\end{thebibliography}
\end{document}